\documentclass[12pt,preprint,trackchanges]{aastex63}
\usepackage{amsmath}

\usepackage[normalem]{ulem}

\begin{document}

%\title{Constraints on Fast Radio Burst Scattering from the First CHIME/FRB Catalog}
%\title{Dispersion and Scattering of Fast Radio Bursts: Constraints from the first CHIME/FRB Catalog}
\title{Modeling Fast Radio Burst Dispersion and Scattering Properties in the First CHIME/FRB Catalog}

\author[0000-0002-3426-7606]{P.~Chawla}
\affiliation{Department of Physics, McGill University, 3600 rue University, Montr\'eal, QC H3A 2T8, Canada}
\affiliation{McGill Space Institute, McGill University, 3550 rue University, Montr\'eal, QC H3A 2A7, Canada}
\affiliation{Anton Pannekoek Institute for Astronomy, University of Amsterdam, Science Park 904, 1098 XH Amsterdam, The Netherlands}

\author[0000-0001-9345-0307]{V.~M.~Kaspi}
\affiliation{Department of Physics, McGill University, 3600 rue University, Montr\'eal, QC H3A 2T8, Canada}
\affiliation{McGill Space Institute, McGill University, 3550 rue University, Montr\'eal, QC H3A 2A7, Canada}

\author[0000-0001-5799-9714]{S.~M.~Ransom}
\affiliation{National Radio Astronomy Observatory, 520 Edgemont Rd, Charlottesville, VA 22903, USA}

\author[0000-0002-3615-3514]{M.~Bhardwaj}
\affiliation{Department of Physics, McGill University, 3600 rue University, Montr\'eal, QC H3A 2T8, Canada}
\affiliation{McGill Space Institute, McGill University, 3550 rue University, Montr\'eal, QC H3A 2A7, Canada}

\author[0000-0001-8537-9299]{P.~J.~Boyle}
\affiliation{Department of Physics, McGill University, 3600 rue University, Montr\'eal, QC H3A 2T8, Canada}
\affiliation{McGill Space Institute, McGill University, 3550 rue University, Montr\'eal, QC H3A 2A7, Canada}

\author[0000-0002-2349-3341]{D.~Breitman}
\affiliation{Department of Physics, University of Toronto, 60 St.~George Street, Toronto, ON M5S 1A7, Canada}
\affiliation{Dunlap Institute for Astronomy \& Astrophysics, University of Toronto, 50 St.~George Street, Toronto, ON M5S 3H4, Canada}
\affiliation{David A.~Dunlap Department of Astronomy \& Astrophysics, University of Toronto, 50 St.~George Street, Toronto, ON M5S 3H4, Canada}

\author[0000-0003-2047-5276]{T.~Cassanelli}
\affiliation{Dunlap Institute for Astronomy \& Astrophysics, University of Toronto, 50 St.~George Street, Toronto, ON M5S 3H4, Canada}
\affiliation{David A.~Dunlap Department of Astronomy \& Astrophysics, University of Toronto, 50 St.~George Street, Toronto, ON M5S 3H4, Canada}

\author[0000-0003-2319-9676]{D.~Cubranic}
\affiliation{Department of Physics and Astronomy, University of British Columbia, 6224 Agricultural Road, Vancouver, BC V6T 1Z1 Canada}

\author[0000-0003-4098-5222]{F.~Q.~Dong}
\affiliation{Department of Physics and Astronomy, University of British Columbia, 6224 Agricultural Road, Vancouver, BC V6T 1Z1 Canada}

\author[0000-0001-8384-5049]{E.~Fonseca}
\affiliation{Department of Physics, McGill University, 3600 rue University, Montr\'eal, QC H3A 2T8, Canada}
\affiliation{McGill Space Institute, McGill University, 3550 rue University, Montr\'eal, QC H3A 2A7, Canada}

\author[0000-0002-3382-9558]{B.~M.~Gaensler}
\affiliation{Dunlap Institute for Astronomy \& Astrophysics, University of Toronto, 50 St.~George Street, Toronto, ON M5S 3H4, Canada}
\affiliation{David A.~Dunlap Department of Astronomy \& Astrophysics, University of Toronto, 50 St.~George Street, Toronto, ON M5S 3H4, Canada}

\author[0000-0001-5553-9167]{U.~Giri}
\affiliation{Perimeter Institute for Theoretical Physics, 31 Caroline Street N, Waterloo, ON N25 2YL, Canada}
\affiliation{Department of Physics and Astronomy, University of Waterloo, Waterloo, ON N2L 3G1, Canada}

\author[0000-0003-3059-6223]{A.~Josephy}
\affiliation{Department of Physics, McGill University, 3600 rue University, Montr\'eal, QC H3A 2T8, Canada}
\affiliation{McGill Space Institute, McGill University, 3550 rue University, Montr\'eal, QC H3A 2A7, Canada}

\author[0000-0003-4810-7803]{J.~F.~Kaczmarek}
\affiliation{Dominion Radio Astrophysical Observatory, Herzberg Research Centre for Astronomy and Astrophysics, National Research Council Canada, PO Box 248, Penticton, BC V2A 6J9, Canada}

\author[0000-0002-4209-7408]{C.~Leung}
\affiliation{MIT Kavli Institute for Astrophysics and Space Research, Massachusetts Institute of Technology, 77 Massachusetts Ave, Cambridge, MA 02139, USA}
\affiliation{Department of Physics, Massachusetts Institute of Technology, 77 Massachusetts Ave, Cambridge, MA 02139, USA}

\author[0000-0002-4279-6946]{K.~W.~Masui}
\affiliation{MIT Kavli Institute for Astrophysics and Space Research, Massachusetts Institute of Technology, 77 Massachusetts Ave, Cambridge, MA 02139, USA}
\affiliation{Department of Physics, Massachusetts Institute of Technology, 77 Massachusetts Ave, Cambridge, MA 02139, USA}

\author[0000-0002-0772-9326]{J.~Mena-Parra}
\affiliation{MIT Kavli Institute for Astrophysics and Space Research, Massachusetts Institute of Technology, 77 Massachusetts Ave, Cambridge, MA 02139, USA}

\author[0000-0003-2095-0380]{M.~Merryfield}
\affiliation{Department of Physics, McGill University, 3600 rue University, Montr\'eal, QC H3A 2T8, Canada}
\affiliation{McGill Space Institute, McGill University, 3550 rue University, Montr\'eal, QC H3A 2A7, Canada}

\author[0000-0002-2551-7554]{D.~Michilli}
\affiliation{Department of Physics, McGill University, 3600 rue University, Montr\'eal, QC H3A 2T8, Canada}
\affiliation{McGill Space Institute, McGill University, 3550 rue University, Montr\'eal, QC H3A 2A7, Canada}

\author[0000-0002-3777-7791]{M.~M\"unchmeyer}
\affiliation{Department of Physics, University of Wisconsin-Madison, 1150 University Ave, Madison, WI 53706, USA}

\author[0000-0002-3616-5160]{C.~Ng}
\affiliation{Dunlap Institute for Astronomy \& Astrophysics, University of Toronto, 50 St.~George Street, Toronto, ON M5S 3H4, Canada}

\author[0000-0003-3367-1073]{C.~Patel}
\affiliation{Department of Physics, McGill University, 3600 rue University, Montr\'eal, QC H3A 2T8, Canada}
\affiliation{Dunlap Institute for Astronomy \& Astrophysics, University of Toronto, 50 St.~George Street, Toronto, ON M5S 3H4, Canada}

\author[0000-0002-8912-0732]{A.~B.~Pearlman}
\altaffiliation{McGill Space Institute~(MSI) Fellow and FRQNT Postdoctoral Fellow.}
\affiliation{Department of Physics, McGill University, 3600 rue University, Montr\'eal, QC H3A 2T8, Canada}
\altaffiliation{FRQNT Postdoctoral Fellow.}
\affiliation{McGill Space Institute, McGill University, 3550 rue University, Montr\'eal, QC H3A 2A7, Canada}

\author[0000-0002-9822-8008]{E.~Petroff}
\altaffiliation{Veni Fellow.}
\affiliation{Department of Physics, McGill University, 3600 rue University, Montr\'eal, QC H3A 2T8, Canada}
\affiliation{McGill Space Institute, McGill University, 3550 rue University, Montr\'eal, QC H3A 2A7, Canada}
\affiliation{Anton Pannekoek Institute for Astronomy, University of Amsterdam, Science Park 904, 1098 XH Amsterdam, The Netherlands}

\author[0000-0002-4795-697X]{Z.~Pleunis}
\affiliation{Department of Physics, McGill University, 3600 rue University, Montr\'eal, QC H3A 2T8, Canada}
\affiliation{McGill Space Institute, McGill University, 3550 rue University, Montr\'eal, QC H3A 2A7, Canada}

\author[0000-0003-1842-6096]{M.~Rahman}
\affiliation{Sidrat Research, PO Box 73527 RPO Wychwood, Toronto, ON M6C 4A7, Canada}

\author[0000-0001-5504-229X]{P.~Sanghavi}
\affiliation{Lane Department of Computer Science and Electrical Engineering, 1220 Evansdale Drive, PO Box 6109  Morgantown, WV 26506, USA}
\affiliation{Center for Gravitational Waves and Cosmology, West Virginia University, Chestnut Ridge Research Building, Morgantown, WV 26505, USA}

\author[0000-0002-6823-2073]{K.~Shin}
\affiliation{MIT Kavli Institute for Astrophysics and Space Research, Massachusetts Institute of Technology, 77 Massachusetts Ave, Cambridge, MA 02139, USA}
\affiliation{Department of Physics, Massachusetts Institute of Technology, 77 Massachusetts Ave, Cambridge, MA 02139, USA}

\author[0000-0002-2088-3125]{K.~M.~Smith}
\affiliation{Perimeter Institute for Theoretical Physics, 31 Caroline Street N, Waterloo, ON N25 2YL, Canada}

\author[0000-0001-9784-8670]{I.~Stairs}
\affiliation{Department of Physics and Astronomy, University of British Columbia, 6224 Agricultural Road, Vancouver, BC V6T 1Z1 Canada}

\author[0000-0003-2548-2926]{S.~P.~Tendulkar}
\affiliation{Department of Astronomy and Astrophysics, Tata Institute of Fundamental Research, Mumbai, 400005, India}
\affiliation{National Centre for Radio Astrophysics, Post Bag 3, Ganeshkhind, Pune, 411007, India}
  
\correspondingauthor{P. Chawla}
\email{pragya.chawla@mail.mcgill.ca}

\begin{abstract}
We present a Monte Carlo-based population synthesis study of fast radio burst (FRB) dispersion and scattering focusing on the first catalog of sources detected with the Canadian Hydrogen Intensity Mapping Experiment Fast Radio Burst (CHIME/FRB) project. We simulate intrinsic properties and propagation effects for a variety of FRB population models and compare the simulated distributions of dispersion measures (DMs) and scattering timescales with the corresponding distributions from the CHIME/FRB catalog. Our simulations confirm the results of previous population studies, which suggested that the interstellar medium of the host galaxy alone (simulated based on the NE2001 model) cannot explain the observed scattering timescales of FRBs. We therefore consider additional sources of scattering, namely, the circumgalactic medium (CGM) of intervening galaxies and the circumburst medium whose properties are modeled based on typical Galactic plane environments. We find that a population of FRBs with scattering contributed by these media is marginally consistent with the CHIME/FRB catalog. In this scenario, our simulations favor a population of FRBs offset from their galaxy centers over a population which is distributed along the
spiral arms. However, if the models proposing the CGM as a source of intense scattering are incorrect, then we conclude that FRBs may inhabit environments with more extreme properties than those inferred for pulsars in the Milky Way. 
\end{abstract}

\section{Introduction} \label{sec:intro}
Fast radio bursts (FRBs) are dispersed transients of $\mu$s to ms duration detectable at radio frequencies (see \citealt{petroff19} and \citealt{cordes19} for reviews). Their dispersion measures (DMs) represent the integrated electron column density along the line of sight. The measured DMs are in excess of those expected from the Milky Way, suggesting that the bursts originate outside our Galaxy. Identification of the host galaxy for 20 FRBs has confirmed that these bursts are located at cosmological distances\footnote{\url{http://frbhosts.org}; accessed in November 2021} (\citealt{heintz20} and references therein). 

The physical origin of FRBs is as yet unknown, with potential progenitors ranging from isolated neutron stars to mergers of compact objects (see \citealt{platts19} for a summary of the proposed models\footnote{\url{http://frbtheorycat.org}}). The recent detection of a ms-duration radio burst from the Galactic magnetar SGR 1935+2154 shows that magnetars produce at least some fraction of the FRB population \citep{bochenek20, chime20}. Additionally, cataclysmic models can be ruled out at least for the FRB sources that have been observed to repeat (see, e.g., \citealt{spitler16,chime19b}). It is still unclear whether repeating FRB sources have  different progenitors from the so-far non-repeating FRBs, but recent studies suggest that some of the observed properties, namely,  burst widths and bandwidths, are significantly different for the two populations \citep{chime19b,fonseca20,pleunis21}.  

FRBs can be used as probes of electron density in the intervening media even without knowledge of their progenitors and emission mechanism. As FRB sightlines encounter baryons occurring in a highly diffuse state in the intergalactic medium (IGM), the measured DMs of localized FRBs have been used to constrain the cosmic baryon density \citep{macquart20}. Variations in DM along FRB sightlines can probe electron density fluctuations in the IGM and constrain the length scales for IGM turbulence \citep{mcquinn14,masui15b,xu20}. Additionally, FRB DMs can be used to study the diffuse ionized gas in the halos of the Milky Way and other Local Group galaxies (see, e.g., \citealt{prochaska19, platts20}). 
 
Another property of FRBs that can be used to study intervening plasma is scattering. Multi-path propagation in inhomogeneous plasma can cause an FRB pulse to be broadened and exhibit an asymmetric scattering tail. For some FRBs located at low Galactic latitudes, such as FRB 20121102A and FRB 20180916B, the Milky Way is inferred to be the dominant contributor to the observed scattering \citep{ocker21}. However, most FRBs have scattering times greater than the Galactic expectation along the line of sight \citep{cordes16,cordes19}. While the contribution of the Milky Way to FRB scattering is well constrained by Galactic models of electron density \citep{cordes02,yao17}, the dominant source of FRB scattering is uncertain.

No correlation has been observed between FRB scattering times and DMs \citep{katz16,cordes16,qiu20}. Since the IGM contributes significantly to FRB DMs \citep{shannon18}, the lack of a correlation between DMs and scattering times suggests that the IGM is not a significant source of scattering. \citet{masui15} also argue against the IGM being the dominant source of scattering for FRB 20110523A and suggest that scattering material is located either in the host galaxy or the circumburst environment. More recently, \citet{chittidi20} and \citet{simha20} concluded that the scattering for FRB 20190608B originates in either the spiral arm of the host galaxy or the circumburst environment and ruled out intervening halo gas as a source of scattering.
%\citet{day20} reported that the two FRBs with the largest scattering timescales in their localized FRB sample reside in the outskirts of their host galaxies. \pc{The observation implies that} the circumburst environment, and not the interstellar medium (ISM) of the host galaxy, is the dominant source of scattering.  

While the aforementioned studies focus on individual FRBs, it is important to ascertain the dominant source of scattering for the population as a whole. If scattering originates in the circumburst medium, then observed scattering timescales can help determine its properties and allow for constraints on progenitor models \citep{chime19a}. If the majority of scattering is found to be caused by halos of intervening galaxies instead, as is suggested by \citet{vedantham19}, then FRB scattering can enable studies of these halos which are poorly understood. Additionally, population studies can also probe the level of plasma turbulence in galaxies that host FRBs \citep{cordes16}.

Population synthesis studies of FRB scattering involve simulating host galaxies and other intervening media to assess whether they can reproduce the observed properties for the population. Using this approach, \citet{chime19a} concluded that FRBs are located in environments with stronger scattering properties than those derived for the Milky Way ISM using the NE2001 model. Furthermore, \citet{hackstein20} found that the scattering timescales for bursts detected with the Parkes (Murriyang) telescope can only be explained by a more turbulent environment than that of SGR 1935+2154. However, these studies were conducted on small FRB samples without absolute calibration of observational biases. 

Here we report on a population synthesis study to interpret the scattering properties of the first catalog of bursts detected with the Canadian Hydrogen Intensity Mapping Experiment Fast Radio Burst project (hereafter CHIME/FRB; \citealt{chime18}). The catalog includes 535 bursts and is the largest sample of FRBs detected using a single detection pipeline with a well-characterized selection function \citep{chime21}. 

The paper's outline is as follows. In \S \ref{sec:observations}, we discuss our observational inputs which include the dispersion and scattering properties of the CHIME/FRB catalog. Simulated models for intrinsic FRB properties and propagation effects are described in \S \ref{sec:intrinsic} and \S \ref{sec:prop}, respectively. The procedure for the simulations is detailed in \S \ref{sec:procedure} while results are presented and discussed in \S \ref{sec:results}. We summarize our conclusions in \S \ref{sec:conclusion}. We adopt the Planck cosmological parameters \citep{planck16} throughout this paper.

\section{Observations} \label{sec:observations}
The first CHIME/FRB catalog (hereafter ``Catalog 1") contains 474 so-far non-repeating sources and 61 bursts from 18 previously reported repeating sources observed in the interval from 2018 July 25 to 2019 July 2 \citep{chime21}. These bursts were detected in the frequency range of 400--800 MHz by searching 1024 total intensity (Stokes I) sky beams with a 0.983-ms time resolution. 

The burst-fitting process is described in detail by \citet{chime21}. In summary, intensity data for the beam in which the burst was detected with the highest S/N is processed using a least-squares fitting routine, \texttt{fitburst}\footnote{The fitburst code has not yet been made public, but the underlying model and likelihood are the same as that used
by \citet{masui15}, whose code is public.}. The two-dimensional dynamic spectra of the bursts is modeled as the product of two terms describing the  time-independent spectral energy distribution (SED) and the temporal shape, respectively. The SED is modeled as a power-law function with an extra “running” term $r$ in the exponent. The intensity at  frequency $\nu$, $I(\nu) = (\nu/\nu_0)^{-\alpha + r \textrm{ln}(\nu/\nu_0)}$, where $\alpha$ is the spectral index and $\nu_0$ is the reference frequency (set to be equal to 400.1953125 MHz). The temporal shape is described by a Gaussian function convolved with an exponential scattering tail, the timescale for which has a power-law index of $-4$ for the frequency dependence. 

The routine fits for intrinsic width, DM, scattering timescale, burst amplitude, time of arrival, spectral index ($\alpha$) and spectral running ($r$). A second model assuming no scattering is also fit to each burst and $\chi^2$ values for the two models are compared to assess whether the detection of scattering is statistically significant. If the statistical significance is less than 3$\sigma$, the scattering time is reported as an upper limit approximately equal to the intrinsic width. This criterion results in scattering times for 257 bursts being reported as upper limits.

The observed distribution of best-fit scattering times for the 535 bursts in the sample is not equivalent to the intrinsic distribution for the FRB population and requires correction for selection biases. As a start, bursts susceptible to unquantifiable selection effects are removed from the scattering time distribution. Detailed criteria for exclusion are presented by \citet{chime21},  but we note that these exclude bursts with DMs $<$ 100 pc cm$^{-3}$. Bursts with DMs $<1.5$ times the maximum of the Galactic DM estimates based on the NE2001 \citep{cordes02} and YMW16 \citep{yao17} models are also excluded. 

Additionally, we only include the first detection from each repeating FRB source. Subsequent bursts are excluded to ensure that repeating FRBs are not overrepresented in the scattering time distribution for the full population. In doing so, we assume that the scattering timescales do not change between detection of repeat bursts. We find that this assumption is valid for all but one of the repeating sources in our sample i.e. scattering times for repeat bursts are consistent with each other at the $3\sigma$ level.
%scattering material in the immediate vicinity of the source (see, e.g., \citealt{mckee18})
%Such behavior has been observed for the Crab Pulsar with the scattering timescale variations being attributed to discrete structures within the Crab Nebula. 

The remaining sample includes 292 bursts from both repeating and non-repeating FRB sources. The distribution of scattering times for the surviving sample is then corrected for selection biases, which can be introduced by the telescope beam, RFI environment, gain calibration and signal-classification techniques. These biases are determined by injection of simulated signals into the real-time detection pipeline \citep{chime21}. In summary, the fluence, DM, pulse width, scattering time, spectral index and running of the simulated signals are sampled from the plausible ranges of these parameters. The detection probabilities determined for the injected signals are used to ascertain the instrument selection function for different burst properties.

We use the resulting selection-corrected distribution of scattering timescales (see Figure 17 in \citealt{chime21}) as an observable to which we compare our simulations. Since dispersion and scattering properties for a given medium are correlated and cannot be studied in isolation, the selection-corrected DM distribution is also used in our analysis. Our population synthesis study tests different FRB population models based on whether they can reproduce this joint distribution of DM and scattering.

The population models that we test assume different spatial distributions and host galaxy
types for the simulated FRBs. Additionally, some of the models invoke scattering originating
in the circumburst medium and/or in the halos of intervening galaxies. For each population
model, we first simulate properties intrinsic to the source, namely, burst energy, redshift, width and
sky location. The prescriptions that we use for simulating these properties are described
in \S\ref{sec:intrinsic}. We then simulate the DM and scattering contributions of different intervening
media, prescriptions for which are detailed in \S\ref{sec:prop}.

Using burst energies and redshifts, we calculate burst fluences in order to determine which of the simulated FRBs are detectable with the CHIME/FRB system. The DM and scattering distributions of the simulated detectable bursts are then compared
with the corresponding distributions in Catalog 1. We do not compare the simulated fluence distribution with the measured fluences in Catalog 1 as the selection-corrected distributions for DM and scattering are derived assuming these properties are uncorrelated with fluence.

\section{Modeling Intrinsic Properties}\label{sec:intrinsic}

\subsection{Redshift}\label{sec:redshifts}
The intrinsic distribution of FRB redshifts is not well-characterized due to the small number of observed FRBs with host galaxy associations \citep{heintz20}.
In the absence of knowledge about the distribution, we test two different models for the variation of the number density of FRBs with redshift. The first model 
assumes a constant comoving number density with the probability of detecting an FRB at a redshift $z$, 
\begin{equation}\label{eq:comoving}
    P(z) \propto \frac{1}{1+z} \frac{dV_C}{d\Omega dz},
\end{equation} 
where $dV_C/(d\Omega dz$) is the differential comoving volume per unit solid angle per unit redshift. The factor of $(1+z)$ corrects the occurence rate of FRBs in each redshift interval $dz$ for time dilation due to cosmic expansion. 
%i.e. twice the number of events would be detected each year in the observer's frame from a unit comoving volume at $z = 0$ as compared to a unit comoving volume at $z = 1$ as 0.5 years will pass in the rest frame of the source at $z = 1$.
\begin{figure}[h]
\centering
    \includegraphics[scale=0.65]{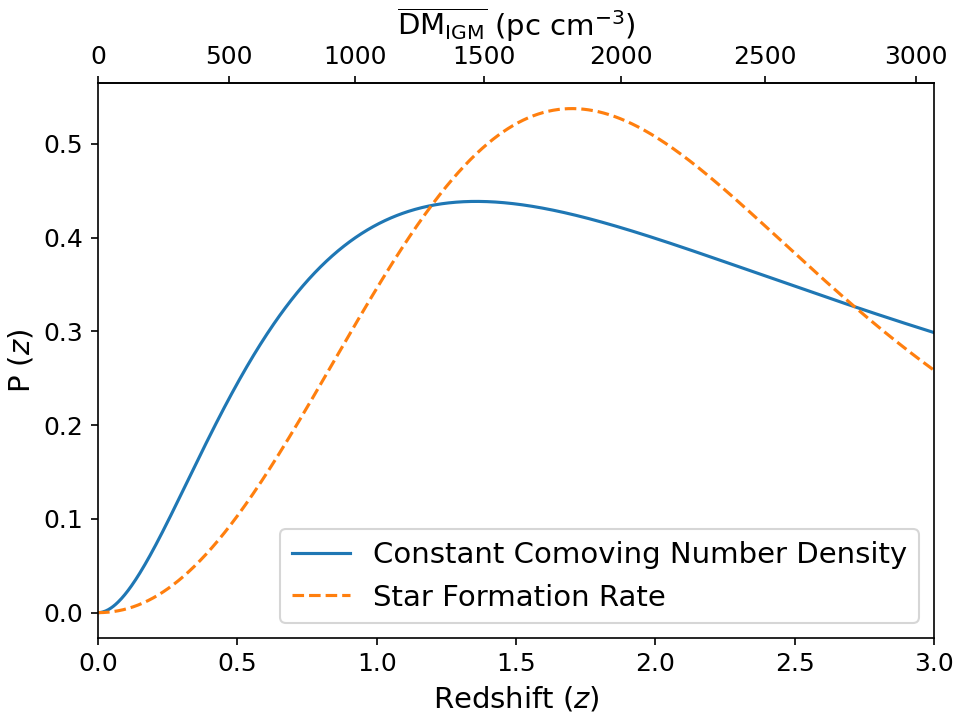}
    \caption{Probability density function of source redshift for two models of the FRB population. The two distributions correspond to comoving number density of FRBs being constant or following the star formation rate (see \S\ref{sec:redshifts}). The top x-axis translates redshift to the mean DM contribution of the intergalactic medium (see \S\ref{sec:igm}). In determining DM$_\mathrm{IGM}$, we assume that the hydrogen and helium in the IGM are fully ionized.}
    \label{fig:redshifts}
\end{figure}
The second model assumes that the number density follows the star-formation rate, motivated by emission models which suggest that FRBs could originate from young stars (see, e.g., \citealt{connor16,margalit18}). For this model, the probability of detecting an FRB at a redshift $z$ is given by
\begin{equation}
    P(z) \propto \frac{(1+z)^{2.7}}{1 + [(1+z)/2.9]^{5.6}} \frac{1}{1+z} \frac{dV_C}{d\Omega dz},
\end{equation} 
where the additional pre-factor describes the variation in the comoving star-formation rate with redshift \citep{madau14}. The probability density functions for the two models are plotted in Figure \ref{fig:redshifts}.

We simulate FRBs up to a maximum redshift of $z = 3$ for both models. This choice is informed by the maximum possible redshift calculated using the \citet{macquart20} relation for the highest-DM event in Catalog 1. The event, FRB 20180906B, has a measured extragalactic DM of 3015 pc cm$^{-3}$, which translates to a maximum redshift of 2.95. The maximum redshift of 3 is a conservative assumption as approximately half of the CHIME-detected FRBs are at redshifts $<0.5$ \citep{chime21}. 
%FRBs with redshifts greater than the assumed maximum would not be bright enough to be detected by the CHIME/FRB system

\subsection{Energy}\label{sec:energy}
We sample the energy emitted by the simulated bursts in the CHIME band according to the Schechter function \citep{schechter76}. While a power-law model is widely used to describe the energy distribution of several extragalactic high-energy transients (see, e.g., \citealt{sun15}), we use the Schechter function as it adds an exponential cutoff to the power-law model. Under this assumption, the differential energy distribution for the FRB population is given by,
\begin{equation}\label{eq:schechter}
    \frac{dN}{dE} \propto E^{-\gamma} \ \mathrm{exp}\bigg(-\frac{E}{E_{\mathrm{max}}}\bigg).
\end{equation}
The exponential cutoff above a maximum energy, $E_{\mathrm{max}}$, ensures that the total energy emitted by the FRB population
($\int_0^\infty E \ (dN/dE) \ dE$) does not diverge for $\gamma \leq 2$. To reduce the computational cost incurred in simulating progressively larger number of FRBs with low energies, we also assume a minimum energy, $E_{\mathrm{min}}$. While we allow the power-law index of the distribution, $\gamma$, to vary for each population model being tested (see \S \ref{sec:procedure}), the minimum and maximum burst energies are held fixed. 

We choose the values of $E_{\mathrm{min}}$ and $E_{\mathrm{max}}$ by studying the inferred energy distribution of events included in Catalog 1. The isotropic-equivalent energy of each event is inferred using the expression derived by \citet{macquart18},
\begin{equation}\label{eq:energy}
    E = \frac{4\pi D_L^2 F \Delta \nu}{(1+z)^{2-\alpha}},
\end{equation}
where $D_L$ is the luminosity distance to the FRB source, $F$ is the burst fluence, $\Delta \nu$ is the observing bandwidth and $\alpha$ is the spectral index ($F_\nu \propto \nu^\alpha$). We set $\alpha=0$ to be consistent with the calibration technique used to measure fluences for the Catalog 1 events. Calibration is performed on a band-averaged time series for each burst and thus involves the implicit assumption of $\alpha=0$ \citep{chime21}. 

\begin{figure}[h!]
\centering
    \includegraphics[scale=0.65]{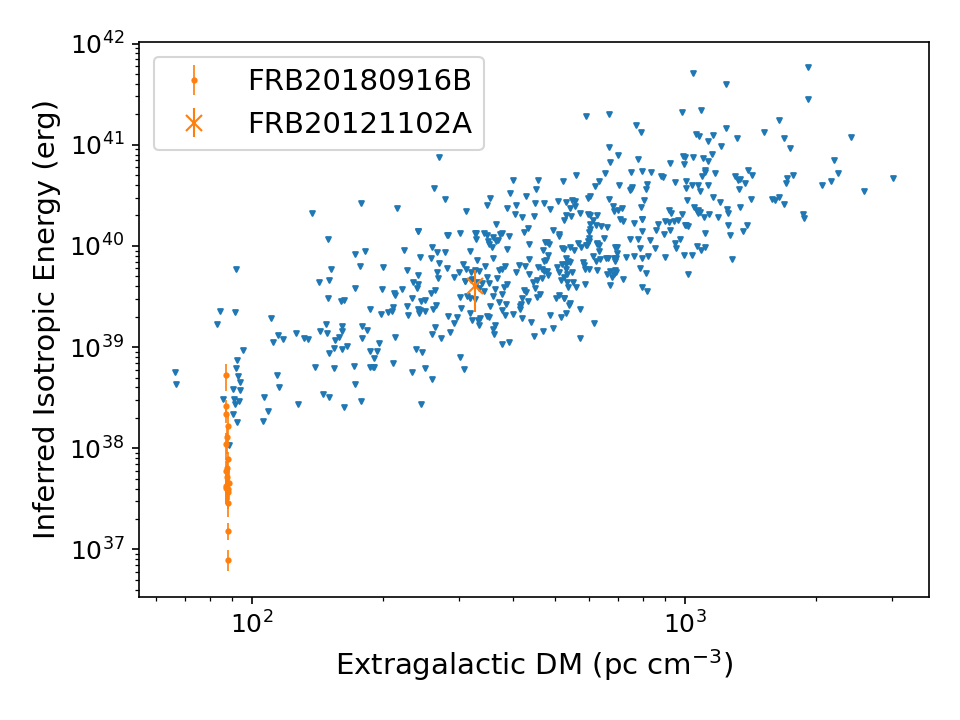}
    \caption{Inferred isotropic-equivalent energy for Catalog 1 events plotted as a function of their extragalactic DM. Events with measured redshifts are repeaters and are plotted in orange with the error bars corresponding to 1$\sigma$ uncertainties propagated from the fluence measurements. Events plotted in blue do not have corresponding redshift measurements. Their energy is estimated using upper limits on the source redshift and burst fluence (see \S\ref{sec:energy}).}
    \label{fig:energycutoff}
\end{figure}

Equation \ref{eq:energy} requires the redshift to be known, which is the case for only two sources in Catalog 1, FRB20121102A and FRB20180916B \citep{tendulkar17,josephy19,marcote20}. For these sources, we compute burst energies with the associated uncertainties being propagated from the fluence measurements. For all other sources, i.e., those with undetermined redshifts, we compute a maximum energy by ascribing the excess DM to the IGM and calculating a maximum possible redshift using the Macquart relation \citep{macquart20}. 
The inferred burst energies are plotted in Figure \ref{fig:energycutoff}. 

Based on the inferred energy distribution, we adopt a value of $10^{42}$ erg for $E_{\mathrm{max}}$. We note that the fluence measurements in Catalog 1 are biased low as they are derived assuming the burst is detected along the meridian, which is the most sensitive location along the transit path. This effect is not accounted for in the fluence uncertainties, thereby causing the maximum inferred energy to not be a true upper bound. Therefore, we re-examine the assumed value of $E_{\mathrm{max}}$ in \S\ref{sec:varyparameters}. We choose the value of $E_{\mathrm{min}}$ based on the inferred energies for the two sources with measured redshifts. While the lowest inferred energy is 7$\times 10^{36}$ erg, we adopt a more conservative value of $10^{36}$ erg for $E_{\mathrm{min}}$. 

\subsection{Sky Location}
We simulate sky locations in order to determine the contribution of the Milky Way to the burst DMs and scattering timescales. Burst locations are simulated based on the exposure of the CHIME/FRB system during the observing duration for Catalog 1 (2018 July 25 to 2019 July 1). Since the sky exposure does not vary significantly with right ascension \citep{chime21}, we sample the burst right ascensions from a uniform distribution. However, the exposure varies significantly with declination ($\delta$), 
with sky locations at $\delta > +70^{\circ}$ transiting through the field of view twice per day. 

\begin{figure}[h!]
\centering
    \includegraphics[scale=0.65]{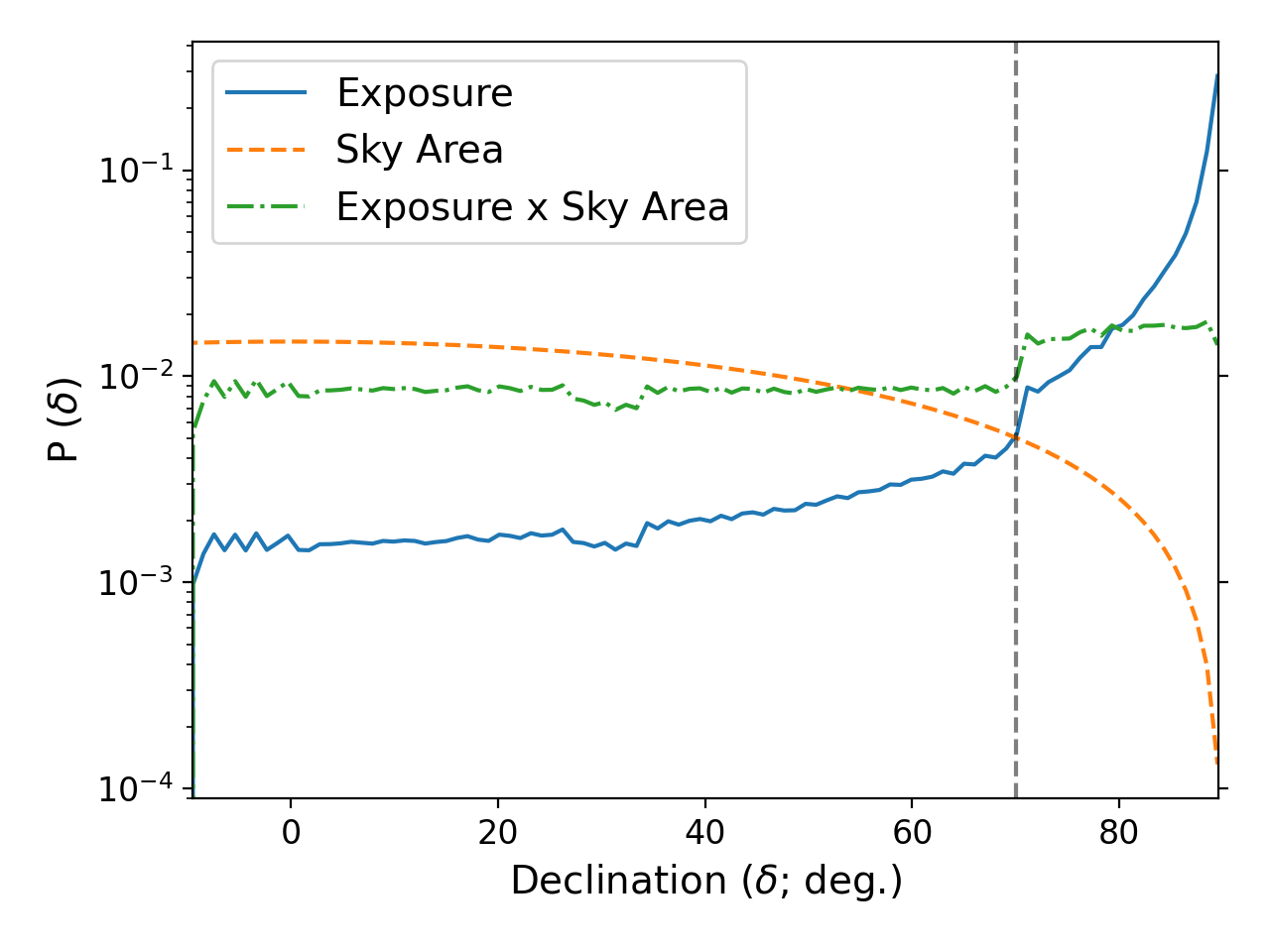}
    \caption{Probability density function (PDF) of burst declinations sampled based on the sky exposure of the CHIME/FRB system. The exposure marginalized over right ascension is plotted in blue while the orange curve shows variation of sky area with declination. The simulated FRB distribution follows the product of these two PDFs represented by the green curve. The dashed grey line marks the declination at which the exposure doubles due to circumpolar sources being visible in two transits.}
    \label{fig:locations}
\end{figure}

In order to characterize the variation in the total exposure of the CHIME/FRB system as a function of declination, we first co-add exposure maps for the two transits provided by \citet{chime21}. While the native resolution of this combined map is 0.7 sq. arcminutes, we downsample the map to a resolution of 1 sq. deg. in order to smooth over the small-scale features such as zero exposure in the gaps between synthesized beams.
We then marginalize the resulting map over right ascension to obtain the exposure as a function of declination. Multiplying this function with a cosine function, which describes the variation in sky area with declination, provides us with a probability distribution. We sample burst declinations in accordance with this probability distribution, which is shown in Figure \ref{fig:locations}. We note that we do not consider variation in sensitivity with declination or across beams and transits while simulating burst locations, as the selection-corrected distributions of observed properties, which the simulated distributions will be compared to, account for bursts detected in locations with different sensitivities.

%{The reduced mean exposure for sources with declinations between 27 and 34 is due to a time-limited failure of one of the four CPU nodes designated to process data for this declination range.}

\section{Modeling Propagation Effects}\label{sec:prop}
We generate the DMs and scattering timescales of simulated FRBs by modeling the different media through which FRB signals propagate. The DM of each burst is equal to the integrated electron density ($n_e$) along the line of sight and is modeled as, 
\begin{equation}\label{eq:dmtotal}
    \textrm{DM}_\mathrm{total} = \int^D_0 n_e ds \ = \   \textrm{DM}_\mathrm{MW} + \textrm{DM}_\mathrm{IGM} + \frac{\textrm{DM}_\mathrm{Host}}{(1+z)} + \frac{\textrm{DM}_\mathrm{Local}}{(1+z)}, 
\end{equation}
where the subscripts `MW', `IGM', `Host' and `Local' stand for the Milky Way, the intergalactic medium, host galaxy and local environment of the FRB source, respectively. Assumed models of electron density for each of these media are described in the following subsections. The contribution of the halos of intervening galaxies is included in DM$_\mathrm{IGM}$ (see \S\ref{sec:igm}). The DM contribution of the host galaxy and the circumburst environment are both reduced by a factor of (1+$z$) to account for the combined effect of time dilation of the dispersive delay and photon redshift \citep{ioka03}. 

We simulate the phenomenon of scattering using the prescription provided by \citet{cordes02}. They model electron density fluctuations ($\delta n_e$) in the intervening inhomogeneous plasma as having a power-law wavenumber ($q$) spectrum,  
\begin{equation}
    P_{\delta n_e}(q) = C_n^2 q^{-\beta},
\end{equation}
where $C_n^2$ is the spectral coefficient and denotes the level of turbulence. The inner and outer scale of these fluctuations are denoted by $l_i = 2\pi/q_i$ and $l_o = 2\pi/q_o$, respectively, and $\beta$ is set to be equal to $11/3$ for a Kolmogorov wavenumber spectrum. Based on this prescription, we can evaluate the scattering time ($\tau$) using the following expression \citep{blandford85,cordes16},
\begin{align}\label{eq:scat}
    \tau &= \frac{1}{2c} \int_0^{D_A} ds \ \eta(s) s ( 1 - s/D_A )\\
         &= \frac{3 \Gamma(7/6) \lambda^4 r_e^2  q_i^{1/3} }{2c} \int_0^{D_A} ds \ s ( 1 - s/D_A ) \ C_n^2. 
\end{align}
Here $\eta$ is the mean-square scattering angle per unit distance, which we express in Equation 8 in terms of $C_n^2$ following Appendix B of \citet{cordes16}. Additionally, $c$ is the speed of light, $\Gamma$ is the gamma function, $\lambda$ is the observing wavelength (corresponding to a frequency of 600 MHz), $r_e$ is the classical electron radius and $D_A$ is the angular diameter distance to the source. Following \citet{cordes16}, we set the inner scale, $q_i$ to be equal to 10$^3$ km. The finite wavenumber cutoff associated with the inner scale results in a power-law index of $-4$ for the frequency dependence, as opposed to the canonical value of $-4.4$ for a Kolmogorov medium with a negligible inner scale \citep{bhat04}.

The models of electron density assumed for different media can also be used to ascertain scattering measures (SM). The SM integrates the amplitude of the turbulence over the path length that the FRB signal passes through and is estimated using the following equation presented by \citet{cordes02},
\begin{equation}\label{eq:sm}
    \textrm{SM} = 6 \int_0^{D_A} ds \ (s/D_A) \ (1-s/D_A) \ C_n^2.
\end{equation}
The spectral coefficient $C_n^2$ is related to the electron density ($n_e$) in the intervening medium as,
\begin{equation}
    C_n^2 = C_\mathrm{SM} F n_e^2,
\end{equation}
where $C_\mathrm{SM}$ is a numerical constant derived based on the slope of the wavenumber spectrum and $F$ is the fluctuation parameter which depends on the outer scale, the volume filling factor of ionized clouds and the magnitude of electron density fluctuations ($\delta n_e/n_e$) in the medium.
Analogous to the total DM, we obtain the total SM by adding the contributions of all intervening media,
\begin{equation}
    \textrm{SM}_\mathrm{total} = \textrm{SM}_\mathrm{MW} + \textrm{SM}_\mathrm{IGM} + \textrm{SM}_\mathrm{CGM} + \frac{3\textrm{SM}_\mathrm{Host}}{(1+z)^3} + \frac{6\textrm{SM}_\mathrm{Local}}{(1+z)^3}, 
\end{equation}
where the subscript `CGM' stands for the halos of intervening galaxies (see \S\ref{sec:halos}.) The other subscripts are defined in Equation \ref{eq:dmtotal}. We increase $\textrm{SM}_\mathrm{Host}$ and $\textrm{SM}_\mathrm{Local}$ by a factor of 3 and 6, respectively, to account for the increased broadening caused by plane waves from extragalactic sources as compared to spherical waves from sources within the Milky Way \citep{cordes16}. The factor of $(1+z)^3$ in the terms for scattering media at the source redshift corrects for the dilation of the scattering timescale and the observing frequency being redshifted as compared to the emission frequency \citep{macquart13}. The total SM of each FRB is calculated based on its simulated redshift. The SM is then substituted in Equation 8 to calculate a scattering time after appropriately accounting for the prefactors. 

\subsection{Milky Way}\label{sec:MW}
We estimate the contribution of the Milky Way to the dispersion and scattering measures using the NE2001 model of electron density \citep{cordes02}. The software implementation of the model integrates the electron density up to a galactocentric radius of 50 kpc, which is greater than the size of any modeled components, thereby providing the maximum possible values for DM$_\mathrm{MW}$ and SM$_\mathrm{MW}$.

We do not use the more recent YMW16 model \citep{yao17} as it derives Galactic scattering times using a $\tau-$DM relation instead of modeling electron density fluctuations. While the $\tau-$DM relation that \citet{yao17} use is valid for the level of turbulence in the inner Galaxy, the expected scattering per unit DM varies significantly between different regions of the Galaxy. Therefore, the model overestimates the scattering towards the Galactic anticenter region, as described in \citet{ocker21}.

%As the scattering time depends on a geometrical weighting factor, $s (1-s/D)$, it varies with the location of the scattering material ($s$) and the source (see Equation \ref{eq:scat}). Galactic scattering time evaluated using the YMW16 model needs to be corrected for this weighting factor, which requires an assumption to be made about the location of the scattering material. We refrain from making this assumption as the material is likely distributed along the path length through the Galaxy.

In addition to the components of the ISM included in the NE2001 model, we 
%also 
simulate the DM contributed from the halo of the Milky Way. The DM of the Galactic halo, $\mathrm{DM}_\textrm{MW,Halo}$, is greatly uncertain with estimates spanning $\sim$10--100 pc cm$^{-3}$ \citep{dolag15,prochaska19,keating20}. We initially assume $\mathrm{DM}_\textrm{MW,Halo} = 30$ pc cm$^{-3}$ based on the results of \citet{dolag15}. We examine the effect of this assumption on our results in \S \ref{sec:varyparameters}. We do not simulate the scattering contribution of the halo as it is constrained to be at least an order of magnitude lower than the temporal resolution of the CHIME/FRB system \citep{ocker21}.
%$\tau_\textrm{halo}<0.1$ ms at a frequency of 600 MHz, assuming a power-law frequency dependence of $-4$

\subsection{Intergalactic Medium}\label{sec:igm}
The contribution of the IGM to the DMs of FRBs can be calculated using the Macquart relation, which relates the source redshift to DM$_\mathrm{IGM}$ \citep{macquart20}. There is significant scatter in this relation due to the inhomogeneity of the IGM with a majority of sightlines intersecting galaxy halos \citep{mcquinn14}. To account for these sightline variations, we sample the DM$_\mathrm{IGM}$ for each simulated FRB from a normal distribution with the mean calculated as per \citet{zheng14},
\begin{equation}\label{eq:igm}
    \overline{\textrm{DM}_\mathrm{IGM}} \cong n_0 f_e D_L [1 + 0.932z + (0.16 \Omega_m - 0.078)z^2 ]^{-0.5} 
\end{equation}
where $n_0$ is the mean number density of nucleons at $z = 0$, $f_e$ is the fraction of baryons in an ionized state and $\Omega_m$ is the matter density in units of the critical density at $z = 0$. The top x-axis in Figure \ref{fig:redshifts} shows the variation in $\textrm{DM}_\mathrm{IGM}$ as a function of redshift.  
The standard deviation of the aforementioned normal distribution %, $\sigma(\textrm{DM}_\mathrm{IGM})$, 
is obtained from cosmological simulations performed by \citet[see Figure 1 of their paper]{mcquinn14} for $z < 1.5$. The standard deviation varies from $\sim50\%$ at $z=0.1$ to $\sim20\%$ at $z=1.5$. For simulated FRBs with $z > 1.5$, we assume a standard deviation of 20\%. 

We do not simulate scatter-broadening caused by the IGM and set SM$_\mathrm{IGM} = 0$. This choice is motivated by the expected scattering time at 600 MHz, derived based on the turbulence injection scale associated with cosmic structure formation, being at least an order of magnitude lower than the sampling time of the CHIME/FRB system \citep{macquart13,zhu18}.

\subsection{Host Galaxy}\label{sec:host}
We simulate FRBs in different types of host galaxies, namely, spiral, elliptical and dwarf galaxies. We adopt different models of electron density for each galaxy type, which are described in the following subsections. Additionally, we assume that the inclination angles of these host galaxies are uniformly distributed, with the probability of having an inclination angle $i$, $P (i) \propto \sin (i)$. For each FRB, we randomly sample an inclination angle in the range, $0 \leq i \leq \pi/2$, based on the aforementioned probability. We do not simulate inclination angles with values between $\pi/2$ and $\pi$ as the assumed electron density models are symmetric about the plane of the galaxy. 

\begin{figure}[h!]
\centering
    \includegraphics[scale=0.65]{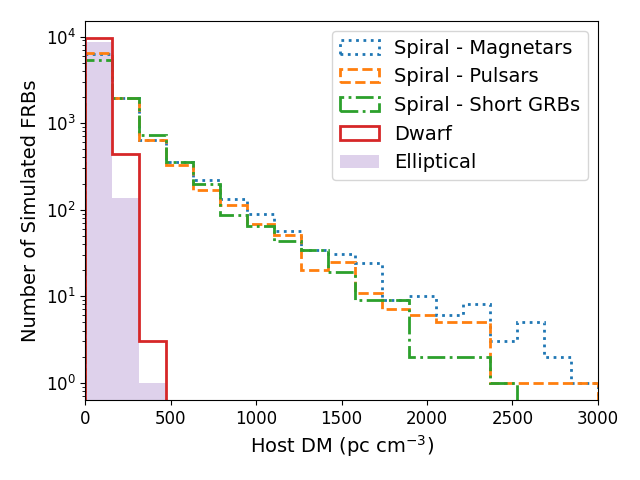}
    \includegraphics[scale=0.65]{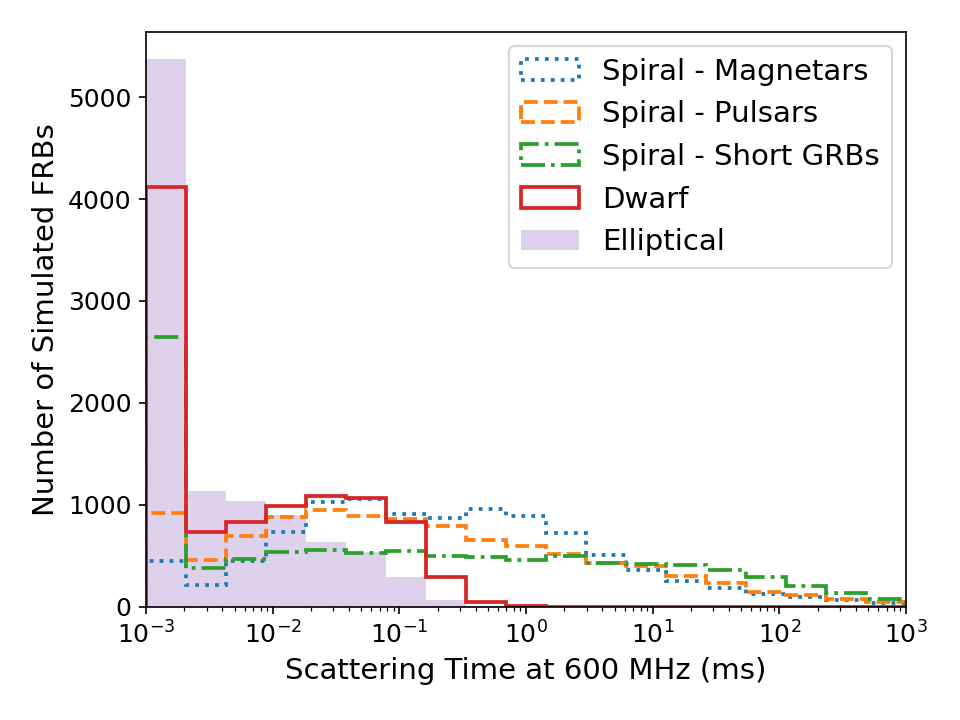}
    \caption{Histogram of the host DMs and scattering times for 10$^4$ simulated FRBs located in different types of host galaxies, namely spiral, dwarf and elliptical galaxies. Spatial distributions of FRBs within the host galaxy are described in \S \ref{sec:host}. Three distributions are considered for spiral galaxies, which emulate the magnetar, pulsar and short GRB populations. For visualization purposes, scattering times $<10^{-3}$ ms are set to be equal to 10$^{-3}$ ms in the bottom panel.}
    \label{fig:hosts}
\end{figure}

We simulate 10$^4$ FRB locations within the host for each galaxy type. The assumed distributions of FRB locations are also described in the following subsections. We then estimate DM$_\textrm{host}$ and SM$_\textrm{host}$ for these 10$^4$ sightlines through the host galaxy by integrating over the electron density distribution from the event location to the near edge of the host galaxy. The resulting distributions of DMs and scattering times are shown in Figure \ref{fig:hosts}. Additionally, we add the DM expected from the halo of the host galaxy to DM$_\textrm{host}$. Analogous to the Milky Way, we assume DM$_\mathrm{Host, Halo}$ = 30 pc cm$^{-3}$. We make this assumption for all galaxy types as the electron distribution in the halos of dwarf and elliptical galaxies is poorly understood. While the known sample of FRB host galaxies is morphologically diverse \citep{heintz20}, we initially test population models where all FRBs are located in host galaxies of the
same type. However, we repeat our analysis for models where FRBs exist in more than one type of host galaxy and present the results in 
%Section 
\S\ref{sec:hostresults}. 

\subsubsection{Spiral Galaxies}\label{sec:spirals}
We assume that all spiral galaxies have electron density distributions similar to the Milky Way and use the prescription from the NE2001 model to simulate the electron density in the thin disk, thick disk, spiral arms and galactic center (see Table 2 of \citealt{cordes02}). In using the NE2001 model to simulate the electron density of the host galaxy, we implicitly assume that the %galaxy size, 
amplitude of turbulence ($C_n^2$) and inner scale of the fluctuations in electron density, $l_i$, are the same as those assumed for the Milky Way by \citet{cordes02}. 
%We examine how our results change for a distribution of host galaxies with different sizes in \S \ref{sec:size}. 

As noted earlier, we do not use the YMW16 model to simulate the host galaxy, as \citet{yao17} use a $\tau-$DM relation (as described in \S \ref{sec:MW}) to obtain the scattering time. As this relation models the expected turbulence in the inner Galaxy, it is only valid for FRBs located in the inner regions of their host galaxy. However, FRBs could be located at large radial distances from their galaxy centers, as discussed below. Their sightlines would not pass through the inner regions of the host, implying that the YMW16 model would overestimate the pulse broadening for such sources.

The distribution of host DMs and scattering times could be influenced by the size distribution of the host galaxies. We assign a radius to each spiral host galaxy by first sampling a galaxy mass and then using the mass-radius relationship provided by \citet{shen03}, 
\begin{equation}\label{eq:radius}
    R \ (\textrm{kpc}) = 0.1 \bigg(\frac{M}{M_\sun}\bigg)^{0.14} \bigg(1 + \frac{M}{4\times10^{10}M_\sun}\bigg)^{0.25}.
\end{equation}
Here $R$ is the galaxy half-light radius and $M$ is the galaxy stellar mass. We sample the galaxy mass using the stellar mass function of star-forming galaxies, located at $z \sim 0.1$, provided in Table 3 of \citet{moustakas13}. In doing so, we assume that the rate of FRB occurence within a galaxy is proportional to its stellar mass. We then derive a radius for each host galaxy based on Equation \ref{eq:radius}, and scale all modeled electron density structures by a factor of ($R/R_\mathrm{MW}$). Here $R_\mathrm{MW}$ is the effective radius of the Milky Way and is set to be equal to 2.5 kpc \citep{vandenbergh99}.

We simulate three spatial distributions of FRBs within a spiral host, modifying any length scales for these distributions by a factor of ($R/R_\textrm{MW}$). The three distributions are motivated by progenitor models involving isolated pulsars \citep{connor16,cordes16b}, isolated magnetars \citep{metzger19} and merging neutron stars \citep{totani13,wang16,margalit19}. For all of these spatial configurations, we sample the source height above the plane of the host galaxy from an exponential distribution. The scale heights of these exponential distributions are set to be equal to those inferred from the observed Galactic population of these sources using population synthesis analyses. We assume a scale height of 30 pc for the magnetar population \citep{olausen14}, 330 pc for the pulsar population \citep{lorimer06} and 800 pc for binary neutron star systems \citep{kiel10}. 

For the pulsar and magnetar models, we simulate FRBs along the spiral arms of the host galaxy. This is motivated by the results of a population synthesis analysis performed by \citet{fgk06} and we use the methodology described in \S3.1.2 of their paper. In summary, the galactocentric radii of simulated FRBs are sampled from a radial distribution for which the surface density peaks at 3 kpc. The azimuthal angle of each FRB is sampled in a way that it falls on the centroids of the spiral arms, locations of which are consistent with those used in the NE2001 model.

For the model in which FRBs are associated with binary neutron star mergers, we simulate the galactocentric radius based on the corresponding distribution observed for short gamma-ray bursts, which exhibit large offsets from the centers of their host galaxies. We sample the radius of the simulated FRBs using the cumulative distribution function (CDF) of host-normalized offsets ($r/R$) provided by \citet{fong13}. 

%Here $r_e$ is the half-light radius of the host galaxy which we assume to be 2.5 kpc, same as that inferred for the Milky Way \citep{vandenbergh99}. 
    
\subsubsection{Elliptical Galaxies}
While we have a prior on the electron density distribution in spiral galaxies owing to extensive modeling of the Milky Way, the corresponding distribution for elliptical galaxies is not well understood. Although some theoretical models provide a functional form for the variation in electron density with radius (see, e.g., \citealt{seta21}), they do not include a prescription for the scattering properties of elliptical galaxies. Therefore, we follow \citet{xu15} in modeling an elliptical galaxy using the thick disk and galactic center components of the NE2001 model, using which we can ascertain both dispersive and scattering properties. 

This approach makes an implicit assumption that the elliptical galaxy is the same size as the Milky Way. However, we reduce the free electron density by a factor of $\sqrt{10}$ in order to account for the lack of ionized gas in elliptical galaxies. This factor is derived based on the H$\alpha$ luminosity for an average elliptical galaxy being an order of magnitude lower than the inferred luminosity for the Milky Way \citep{james04}. As the H$\alpha$ luminosity is a tracer of ionized hydrogen, it is proportional to $\int n_e^2 \ dV$, implying that the free electron density is proportional to the square root of the total H$\alpha$ luminosity emitted by the galaxy.  

We assume a population of FRBs within the galaxy with a scale height and radial distribution the same as that inferred for the short GRB population, described in \S \ref{sec:spirals}. The resulting DM and scattering time distributions are shown in Figure \ref{fig:hosts}. We note that we do not simulate other spatial distributions of FRBs in elliptical galaxies. Another caveat is that reducing the free electron density across the galaxy by a factor based on the H$\alpha$ luminosity assumes that the distribution of ionized hydrogen in elliptical galaxies is similar to that for the Milky Way. This factor could be different if a significant fraction of the H$\alpha$ emission originates in small-scale structures such as \ion{H}{2} regions. Assuming a different spatial distribution for FRBs or correction factor for the H$\alpha$ luminosity is unlikely to significantly change our results. This is because the contribution of an elliptical host galaxy to the total DM and SM will be significantly lower than that of other media, regardless of the assumed parameters.

\subsubsection{Dwarf Galaxies}
We model dwarf galaxies using the prescription for the electron density distribution in the Large Magellanic Cloud (LMC) proposed by \citet{yao17}. Although the LMC is not representative of the full population of dwarf galaxies, observations of radio pulsars have enabled modeling of its electron density distribution. Since similar studies have not been conducted for other galaxies, we use the LMC as a prior for all dwarf galaxies.

Based on DMs observed for 23 pulsars in the LMC, \citet{yao17} model it as a disk with exponentially decreasing electron density away from the plane of the galaxy and a Gaussian distribution in the radial direction. The scale height for the exponential fall-off is 0.8 kpc and the standard deviation for the Gaussian function is 3 kpc. We assume the vertical heights and galactocentric radii of the simulated FRBs to follow an exponential and Gaussian distribution, respectively. The parameters of these distributions are set to be the same as those for the assumed electron density distribution. While integrating over this electron density distribution to calculate the SM of a simulated FRB, we assume the turbulence to be the same as that for the thick disk of the Milky Way since the scattering properties of the LMC are not well understood. This is done by setting $C_n^2$ in Equation \ref{eq:sm} to be the same as that for the thick disk in the NE2001 model. 

We note that we do not implement a distribution of galaxy radii for dwarf or elliptical galaxies as the total DM and scattering time distribution is not expected to change significantly by varying the sizes of these galaxies. This is because the simulated contribution of these galaxy types to the total DMs of simulated FRBs is much lower than the expected IGM contribution. Additionally, the simulated host scattering times, as shown in Figure \ref{fig:hosts}, are lower than the temporal resolution of the CHIME/FRB instrument. 

Two repeating FRB sources, FRB 20121102A and FRB 20190520B, have been localized to star-forming dwarf galaxies \citep{tendulkar17,niu21}. The range of possible values for the host DM of FRB 20121102A ($55 < \textrm{DM}_\mathrm{Host} + \textrm{DM}_\mathrm{Local} < 225$ pc cm$^{-3}$) is consistent with the simulated DMs in this work (see Figure \ref{fig:hosts}). However, the inferred total DM for the host galaxy and local environment of FRB 20190520B is significantly higher (902$^{+88}_{-128}$ pc cm$^{-3}$). The high inferred DM could be consistent with our simulations if the dominant contributor is the local environment instead of the host galaxy ISM. The association of a compact, persistent radio source with FRB 20190520B supports this possibility.

%including its halo and any gas local to the FRB source,

\subsection{Circumburst Environment}\label{sec:local}
The local environment of an FRB source could contribute significantly to its observed DM, as has been suggested for models in which FRBs are produced by young neutron stars (see, e.g., \citealt{connor16,piro16}). However, these works do not provide an estimate of the scatter-broadening caused by high electron density local environments. The NE2001 model of electron density in the Milky Way provides empirical estimates of local DM and SM (hereafter $\textrm{DM}_\mathrm{clump}$ and $\textrm{SM}_\mathrm{clump}$) for $\sim$100 Galactic pulsars \citep{cordes03}. \citet{cordes03} note that most of these clumps 
are not associated with known \ion{H}{2} regions or supernova remnants and hence might not be an adequate measure of the dispersive or scattering properties of such environments. Nevertheless, in the absence of theoretical estimates, we choose to use these empirically derived values as a prior on the scattering properties of \ion{H}{2} regions and supernova remnants. Moreover, the observed dispersion and scattering properties of a few pulsars associated with known \ion{H}{2} regions or supernova remnants \citep{johnston03,rickett09,ocker20} are consistent with the NE2001 estimates for the $\sim$100 pulsars, lending support to our choice of using this sample. 

\begin{figure}[h!]
\centering
    \includegraphics[scale=0.65]{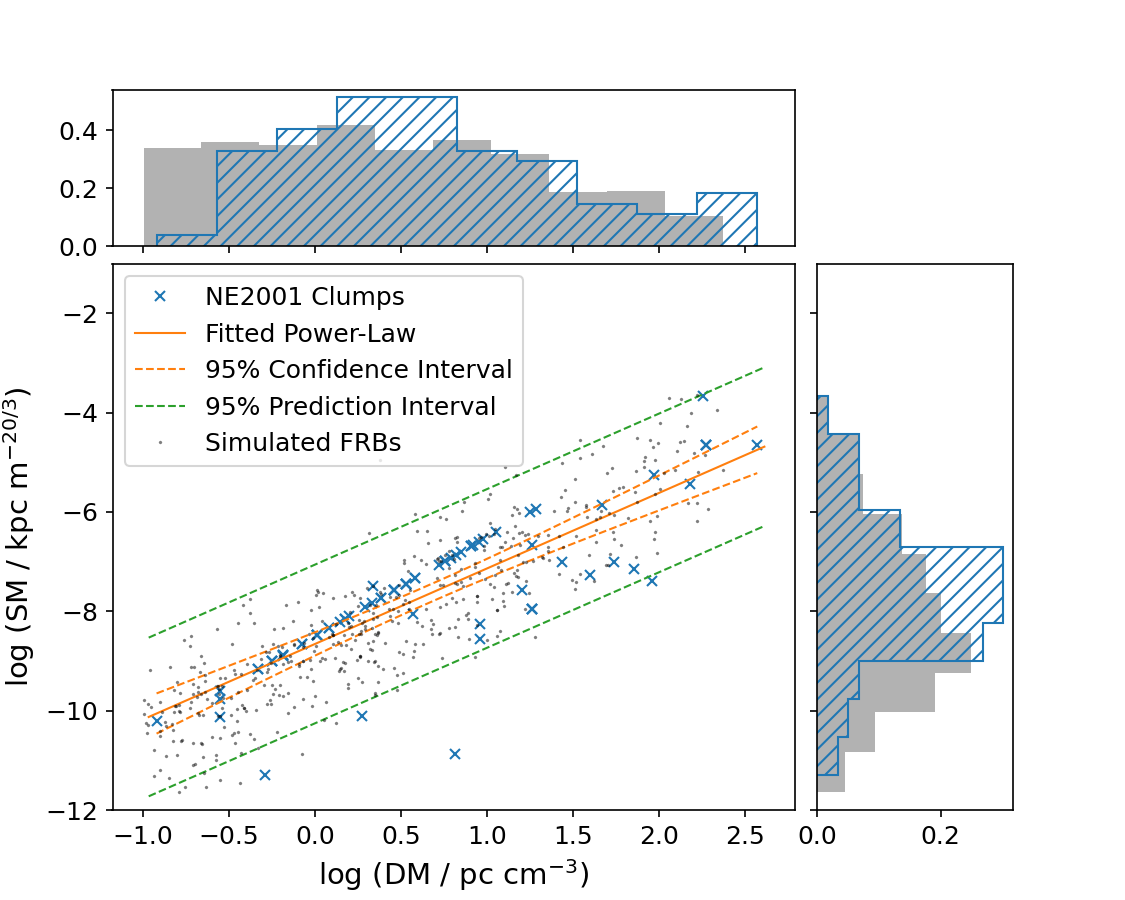}
    \caption{Dispersion and scattering measures for local environments of simulated FRBs in this work and for regions of intense scattering modeled as electron density clumps in the NE2001 model (see \S\ref{sec:local}). Histograms of the distribution of local DMs for the simulated FRBs (in grey) and for the NE2001 clumps (in blue) are plotted in the top panel while the corresponding histograms for the local SMs are shown in the right panel. The DMs of the simulated FRBs are drawn from a log-normal distribution and the SMs are estimated using a power-law model fit to the DM and SM of the NE2001 clumps. The power-law model with the $2\sigma$ confidence interval for the fit is plotted in orange. The scatter in the SMs for the simulated FRBs is dictated by the prediction interval, the bounds for which are shown by the dashed green lines.}
    \label{fig:clump}
\end{figure}

We sample the DM of the local environment, $\textrm{DM}_\mathrm{Local}$, from a log-normal distribution. The parameters of this distribution are based on the distribution of $\textrm{DM}_\mathrm{clump}$ values modeled by \citet{cordes03}. The mean and standard deviation of the underlying normal distribution are 0.18 and 1.09, respectively, and the DM values are generated within 2$\sigma$ of the mean of the normal distribution. Simulated values of $\textrm{DM}_\mathrm{Local}$ range from 0.1 to $\sim$300 pc cm$^{-3}$. The distribution of $\textrm{DM}_\mathrm{clump}$ and $\textrm{DM}_\mathrm{Local}$ are shown in Figure \ref{fig:clump}. 

In order to estimate $\textrm{SM}_\mathrm{Local}$ from the corresponding $\textrm{DM}_\mathrm{Local}$ for each simulated FRB, we first model the relationship between $\textrm{SM}_\mathrm{clump}$ and $\textrm{DM}_\mathrm{clump}$ with a power law using an ordinary least squares regression (see, e.g., \citealt{isobe90}). We then calculate $\textrm{SM}_\mathrm{Local}$ using this power-law model, but allow for scatter in the SM values based on the prediction interval for the least squares regression. The confidence and prediction intervals are shown in Figure \ref{fig:clump}. While the confidence interval represents the uncertainty in the fit, the prediction interval accounts for the scatter in the dependent variable (SM$_\mathrm{clump}$) and indicates the range in which a future observation will lie. Physically, the SM is related to the DM as, 
\begin{equation}
    \textrm{SM}_\mathrm{clump} = C_\mathrm{SM} \bigg(\frac{F_\mathrm{c} \textrm{DM}_\mathrm{clump}^2}{w_\mathrm{c}}\bigg),
\end{equation}
where $F_\mathrm{c}$ is the fluctuation parameter for the clump and $w_\mathrm{c}$ is the clump radius. The SM values simulated within the prediction interval thus sample the range of fluctuation parameters and widths inferred for the local environments of Galactic pulsars.

We test another configuration in which the circumburst environment has more extreme properties than those assumed above. This is done by setting a minimum threshold for $\textrm{DM}_\mathrm{Local}$. While sampling local DMs for this configuration, DMs lower than the threshold are rejected and redrawn. We initially assume a threshold of 10 pc cm$^{-3}$ but examine the effect of increasing the threshold in \S \ref{sec:localconstraints}. The scattering measure, $\textrm{SM}_\mathrm{Local}$, is derived using the same method as described above. 

\subsection{Intervening Galaxies}\label{sec:halos}
The line of sight to an FRB could also intersect one or more galaxies. \citet{macquart13} estimated the probability of an FRB sightline passing through the ISM of an intervening galaxy to be $<5\%$ at $z<1.5$. Since the probability of intersection is fairly low, we do not consider the ISM of intervening galaxies as an additional source of dispersion and scattering. However, the probability of a sightline intersecting the halo of an intervening galaxy is significantly higher. \citet{vedantham19} find that the majority of sightlines out to $z\sim1$ encounter a $10^{13}$ M$_\sun$ halo and approximately ten $10^{11}$ M$_\sun$ halos. The number of halos that a sightline intercepts peaks at $z\sim1$ and decreases thereafter. 

While the DM contribution of such halos is accounted in the scatter in the Macquart relation (see \S \ref{sec:igm}), we model the scattering times using the formalism proposed by \citet{vedantham19}. Their model suggests that sub-parsec scale cool ionized gas clumps (T $\sim 10^4$ K) in the circumgalactic media (CGM) of intervening galaxies could contribute to FRB scattering. 

The scattering time in this model depends on the volume fraction of the cool gas clumps ($f_v$), the fraction of baryons in the CGM ($f_\textrm{CGM}$) and the source redshift. Additionally, the scattering time for two sources at the same redshift could vary by several orders of magnitude depending on the number of galaxy halos their sightlines intersect, masses and redshifts of these halos and the impact parameter to the center of the galaxy. \citet{vedantham19} account for the aforementioned variables to derive an expression for the fraction of sightlines (referred to as the areal covering factor) for which the scattering time exceeds any given value of $\tau$ (see Figure 8 of their paper). The differential of the areal covering factor with respect to $\tau$ provides the probability density function for $\tau$. We sample the scattering time of each FRB based on this probability density function evaluated at the simulated source redshift. 

\citet{vedantham19} note that their model is simplistic with many assumptions and that the values of several key parameters such as $f_v$ and $f_\textrm{CGM}$ are 
poorly constrained. As the validity of the model is uncertain, we perform our population synthesis analysis for two scenarios, with and without scattering originating in the CGM of intervening galaxies. For the former 
scenario, we test two configurations corresponding to 30\% and 60\% of baryons in the CGM, i.e., $f_{CGM}$ = 0.3 and 0.6, setting $f_v = 10^{-4}$ for both these configurations. We choose these two configurations as they can explain scattering observed for FRBs at 1.4 GHz, as noted by \citet{vedantham19}. The scattering time is proportional to $(f_\mathrm{CGM})^{3.2}$ and hence varies by an order of magnitude, on average, between the two configurations.

\section{Simulation Procedure}\label{sec:procedure}
We test different population models of FRBs, which are listed in Table 1, using Monte Carlo simulations. For each population model, we start by simulating properties intrinsic to the source, namely, burst energy, redshift, sky location, and pulse width ($w_i$). We sample pulse widths in accordance with the log-normal function fit to the selection-corrected distribution of intrinsic widths for the Catalog 1 sample. Other intrinsic properties are sampled as described in \S \ref{sec:intrinsic}. Burst energies are sampled according to the Schechter function (see Equation \ref{eq:schechter}) for which the power-law index is initially set to be equal to a fiducial value of 2. 

Based on simulated burst energies and redshifts, we calculate band-averaged fluences for the CHIME band using Equation \ref{eq:energy}. A simulated burst is considered to be detectable if its fluence is greater than 0.4 Jy ms, which is the lowest fluence for any burst in the catalog. We choose this threshold knowing that selection effects for the catalog have been adequately characterized above this 
fluence, thereby
allowing a robust comparison of the simulated distributions with the selection-corrected distributions. Previous population synthesis studies (see, e.g., \citealt{caleb16,gardenier19}) implement a fluence threshold dependent on the pulse width, scattering time and DM, determined using the radiometer equation. Since an absolute calibration of selection effects is performed for Catalog 1, we do not use the radiometer equation as an approximation for the sensitivity of the CHIME/FRB system.

We simulate 10$^4$ bursts for each population model, 
%For each population model, we generate as many bursts as are required to match the number of detectable events in our simulations with the total number of observed events. Specifically, this condition ensures that at least 535 bursts in our simulated sample have a fluence greater than 0.4 Jy ms.
the DMs and scattering times for which are sampled using the prescriptions provided in \S \ref{sec:prop}. All DMs include the contribution of the Milky Way, the host galaxy and the IGM. However, the scattering times only include the contribution of the Milky Way and the host galaxy. While the contributions of the Milky Way and the IGM are derived based on simulated sky locations and redshifts, the host galaxy contribution is simulated for 10$^4$ different sightlines (see \S \ref{sec:host}). Contributions of intervening galaxies and local environments are added for some of the tested models. The different models that we test are listed in Table \ref{tab:models}.

In order to compare the simulated DM and scattering distributions with Catalog 1, we perform the same set of cuts on these distributions as in the catalog. This involves excising bursts with DM $< 100$ pc cm$^{-3}$ or DM $< 1.5$ times the maximum of the Galactic DM estimates obtained from the NE2001 and YMW16 models (see \S \ref{sec:observations}). Additionally, we excise all bursts with $\tau_\textrm{600 MHz} >$ 100 ms, as selection effects for the catalog have not been assessed beyond this timescale \citep{chime21}. Since 95\% of Catalog 1 bursts survived these three cuts, we redraw DMs and scattering times for bursts which are rejected until the surviving sample contains at least 95\% of the simulated detectable bursts, i.e. 9500 bursts.

We also apply two corrections to the simulated distributions to emulate the constraints in measuring widths and scattering times for the Catalog 1 sample. The first of these corrections concerns narrow bursts for which intrinsic widths are difficult to discern due to dispersive smearing and scatter broadening. Simulated bursts with $w_i <$ 0.1 ms are set to have widths equal to 0.1 ms, as \citet{chime21} report that intrinsic widths lower than this threshold cannot be robustly measured. The second correction is applied to bursts with low scattering times ($\tau_\textrm{400 MHz} < w_i$), which are difficult to measure using burst-fitting algorithms. For these bursts, we set $\tau_\textrm{600 MHz}$ to be equal to half the intrinsic width, as is done for Catalog 1 \citep{chime21}.
  
For each population model, we first compare the simulated DM distribution  with the selection-corrected catalog. We generate 100 realizations of the DM distribution for both the catalog and the model. The model distribution includes more than 9500 bursts and thus does not vary significantly between realizations. However, we vary the catalog distribution between realizations to account for the uncertainties shown in Figure \ref{fig:optimal}. This is done by randomly sampling the probability density for each DM bin within its 68\% confidence interval. Since the post-cut catalog consists of 292 bursts (see \S \ref{sec:observations}), we sample 292 DM values in accordance with the PDF using the rejection sampling technique (see, e.g., \citealt{flury90}). 

For each realization, we compare the DM distribution for the catalog with the model using a two-sample Kolmogorov-Smirnov (KS) test \citep{massey51}. If the two distributions are inconsistent with each other, with $>3\sigma$ significance, we resample the intrinsic properties, DMs, and scattering timescales for all bursts assuming a different power-law index for the energy distribution. The power-law index is varied in steps of 0.01 until the two DM distributions are found to be consistent. For most population models, we converge on a power-law index for which the simulated DM distribution is consistent with the catalog. We do not proceed with the scattering time comparison for models which fail to converge. 

We compare the scattering time distribution for 100 realizations of the converged models with the catalog using both the KS and the Anderson-Darling (AD) tests \citep{scholz87}. The probability density function for the selection-corrected scattering time distribution for Catalog 1 has fairly large uncertainties particularly longward of 10 ms, as shown in Figure \ref{fig:optimal}. Therefore, for each of the 100 catalog realizations, we randomly sample the probability density in each scattering time bin within its 68\% confidence interval. We sample 292 scattering timescales based on the resulting PDF and compare these scattering times with the simulated sample of $>9500$ bursts. We report the results from these statistical tests in Table \ref{tab:models}. 
%The probability density function for the simulated distribution (with $>$9500 bursts) does not vary significantly between realizations. 

\floattable
\begin{deluxetable}{llccccccccc}
\tablenum{1}
\caption{P-value for the KS test comparing the selection-corrected scattering time distribution for Catalog 1 and the corresponding simulated distribution for different population models. One hundred realizations are simulated for each model and the simulated scattering distribution is compared to the catalog only if the simulated and observed DM distributions are consistent with each other. The mean p-value for the realizations in which this comparison is performed are reported here. The null hypothesis is that the simulated distribution for a population model is identical to the catalog. The p-value for the model for which at least one of the 100 realizations cannot be ruled out with $>3\sigma$ confidence by both KS and AD tests is indicated in bold. The p-value is not available (as indicated by `...') for models for which the simulated DM distributions are found to be inconsistent with Catalog 1 for all trial power-law indices of the burst energy distribution (see \S \ref{sec:procedure}).} \label{tab:models}
\tablewidth{0pt}
\tablehead{
\multicolumn{2}{c}{DM$_\textrm{Local}$ (pc cm$^{-3}$)} & \multicolumn{3}{c}{--}  & \multicolumn{3}{c}{$> 0.1$\tablenotemark{\textrm{a}}}  & \multicolumn{3}{c}{$> 10$\tablenotemark{\textrm{a}} }\\ 
\multicolumn{2}{c}{Intervening Galaxies (f$_\textrm{CGM}$)\tablenotemark{\textrm{b}}} & \colhead{--} &
\colhead{0.3} &
\colhead{0.6} &
\colhead{--} &
\colhead{0.3} &
\colhead{0.6} & \colhead{--} &
\colhead{0.3} &
\colhead{0.6} \\
\cline{1-11} 
\colhead{Redshift Dist.} & \colhead{Host Galaxy} 
}
\startdata
& Elliptical & 10$^{-82}$ & 10$^{-42}$ & 10$^{-17}$ & 10$^{-77}$ & 10$^{-46}$ & 10$^{-15}$ & 10$^{-39}$ & 10$^{-19}$ & 10$^{-8}$\\
Constant & Dwarf & 10$^{-83}$ & 10$^{-53}$ & 10$^{-25}$ & 10$^{-62}$ & 10$^{-35}$ & 10$^{-21}$ & 10$^{-41}$ & 10$^{-15}$ & 10$^{-10}$ \\
Number & Spiral - Short GRBs & 10$^{-41}$ & 10$^{-22}$ & 10$^{-15}$ & 10$^{-32}$ & 10$^{-19}$ & 10$^{-10}$ & 10$^{-17}$ & 10$^{-7}$ & 10$^{-6}$\\
Density & Spiral - Pulsars & 10$^{-53}$ & 10$^{-30}$ & 10$^{-20}$ & 10$^{-37}$ & 10$^{-23}$ & 10$^{-13}$ & 10$^{-24}$ & 10$^{-10}$ & 10$^{-8}$ \\ 
& Spiral - Magnetars & 10$^{-45}$ & 10$^{-36}$ & 10$^{-31}$ & 10$^{-44}$ & 10$^{-20}$ & 10$^{-21}$ & ... & 10$^{-15}$ & 10$^{-11}$\\
\cline{1-11}
& Dwarf & 10$^{-110}$ & 10$^{-44}$ & 10$^{-19}$ & 10$^{-75}$ & 10$^{-31}$ & 10$^{-17}$ & 10$^{-33}$ & 10$^{-20}$ & 10$^{-9}$\\
SFR\tablenotemark{\textrm{c}} & Spiral - Short GRBs & 10$^{-42}$ & 10$^{-20}$ & 10$^{-15}$ & 10$^{-35}$ & 10$^{-18}$ & 10$^{-9}$ & 10$^{-15}$ & 10$^{-8}$ & \textbf{10$^{-4}$}\\
& Spiral - Pulsars & 10$^{-49}$ & 10$^{-26}$ & 10$^{-14}$ & 10$^{-43}$ & 10$^{-23}$ & 10$^{-15}$ & 10$^{-19}$ & 10$^{-10}$ & 10$^{-12}$\\
& Spiral - Magnetars & 10$^{-51}$ & 10$^{-28}$ & 10$^{-19}$ & 10$^{-33}$ & 10$^{-26}$ & 10$^{-12}$ & 10$^{-23}$ & 10$^{-33}$ & ... \\ 
\enddata
\tablenotetext{a}{The DM of the local environment ranges from the minimum reported here to $\sim$300 pc cm$^{-3}$ and is drawn from a log-normal distribution (see \S \ref{sec:local}).}
\tablenotetext{b}{A dash (--) indicates that the scattering contribution of the CGM of intervening galaxies is assumed to be 0.}
\tablenotetext{c}{For the redshift distribution tracing the star formation rate, we do not test the model in which all FRBs are in elliptical galaxies. The model is not physically motivated due to the low rate of star formation in these galaxies.}
\end{deluxetable}

\section{Results and Discussion}\label{sec:results}
We conclude that a model is able to reproduce the catalog if at least one of 100 Monte Carlo realizations of the DM and scattering distributions for the model is found to be consistent with the corresponding distributions for Catalog 1. The criteria for consistency is finding no significant differences between the observed and simulated distributions at $>3\sigma$ confidence level with both KS and AD tests. Among the several models being tested (see Table \ref{tab:models}), we find that only one population model is able to reproduce the observed properties in Catalog 1. The model that we cannot rule out corresponds to an FRB population hosted in spiral galaxies with a spatial distribution within the host resembling that of the short GRB population. The model also includes scattering arising both in the local environment ($\mathrm{DM}_\textrm{Local} > 10$ pc cm$^{-3}$) and in intervening galaxies ($f_\textrm{CGM} = 0.6$). Additionally, the model assumes that the FRB population evolves with redshift in a manner consistent with the star formation rate. 

For this model, the mean p-value is 10$^{-4}$ for the KS test comparing the simulated scattering time distribution with the catalog. While this value is lower than the threshold for 3$\sigma$ confidence, we do not reject the model as we are performing multiple statistical tests using the same data. Testing multiple hypotheses increases the chances of observing a rare event, thereby requiring a lower p-value threshold to claim an inconsistency between the model and the data. The modified threshold is evaluated using the Bonferroni correction \citep{shaffer95} which involves dividing the threshold for $3\sigma$ confidence (0.0027) by the total number of tests performed (81; see Table \ref{tab:models}). The mean p-value for the favored model is greater than the modified threshold (10$^{-5}$) implying that the model is marginally consistent with the observations. None of the other population models have mean p-values greater than this threshold. We investigate why other models were unable to reproduce our observations in \S \ref{sec:localconstraints} and \S \ref{sec:progenitors}.

\begin{figure}[h!]
\centering
    \includegraphics[scale=0.65]{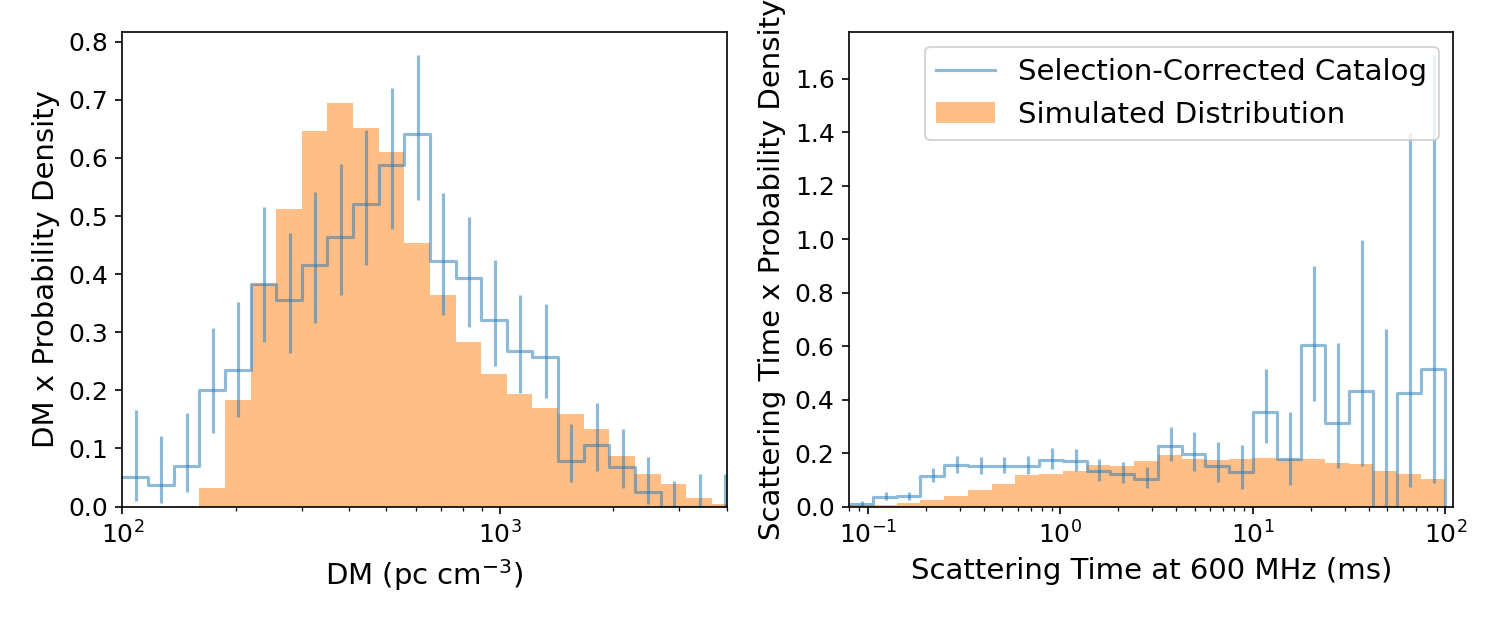}
    \caption{Dispersion measure and scattering time distributions for Catalog 1 and the population model which we cannot rule out. The model assumes FRBs to be spatially distributed like short GRBs in spiral galaxies. The local environment of the FRB source and the CGM of intervening galaxies are the dominant sources of scattering for this model (see Figure \ref{fig:contributions}). The histograms represent the probability density function re-parameterized in terms of the natural logarithm of the two quantities, DM and scattering time. The blue histograms are the catalog data corrected for selection effects with the error bars showing the 68\% Poissonian confidence interval for each bin value. The simulation histograms, plotted in orange, show the average PDF over 100 Monte Carlo realizations of the model. While the model does not adequately reproduce the scattering time distribution, it is found to be marginally consistent based on KS and AD tests and thus cannot be ruled out (see Table \ref{tab:models}). The difference between the probability density function of the simulated and observed scattering distributions for timescales $<1$ ms is not significant as is shown by the corresponding cumulative distribution functions plotted in Figure \ref{fig:cdfoptimal}.}
    \label{fig:optimal}
\end{figure}

\begin{figure}[h!]
\centering
    \includegraphics[scale=0.65]{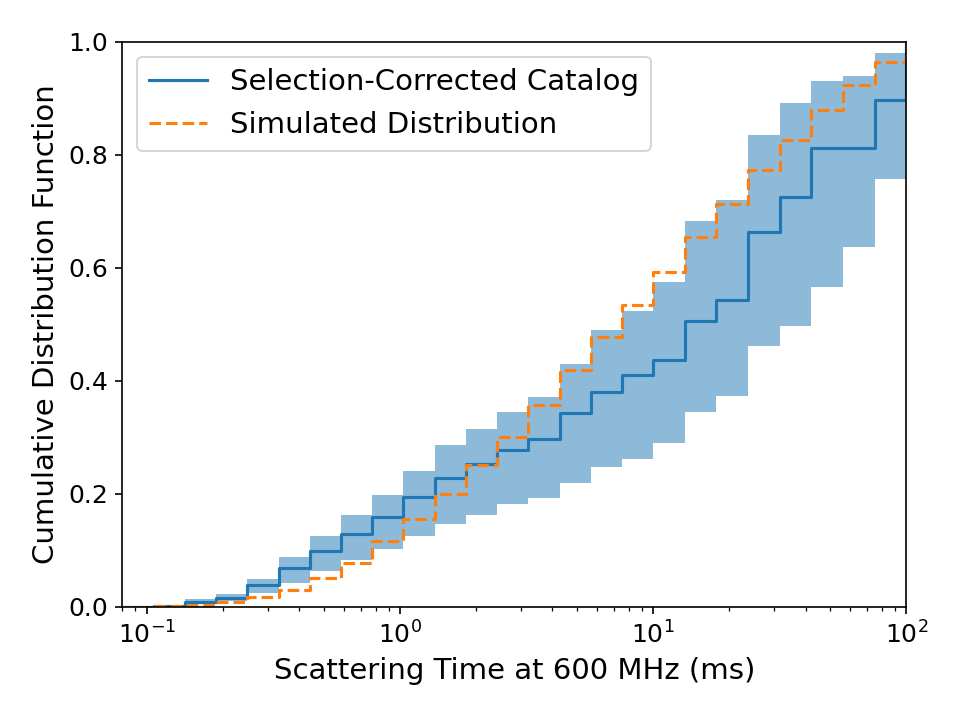}
    \caption{Cumulative distribution function for the scattering times in Catalog 1 and the model which we cannot rule out. The model assumes that FRBs are spatially distributed like short GRBs in spiral galaxies. The model includes scattering originating in the local environment and in the CGM of intervening galaxies. The CDF for the simulated distribution, plotted in orange, is averaged over 100 Monte Carlo realizations of the model. The simulated distribution is found to be marginally consistent with Catalog 1 based on KS and AD tests. The CDF for the selection-corrected catalog is plotted in blue. Uncertainties on the catalog CDF are determined by generating 100 realizations of the probability density function shown in Figure \ref{fig:optimal}. In each realization, the probability density for each scattering time bin is sampled within its 68\% confidence interval.}
    \label{fig:cdfoptimal}
\end{figure}

\begin{figure}[h!]
\centering
    \includegraphics[scale=0.65]{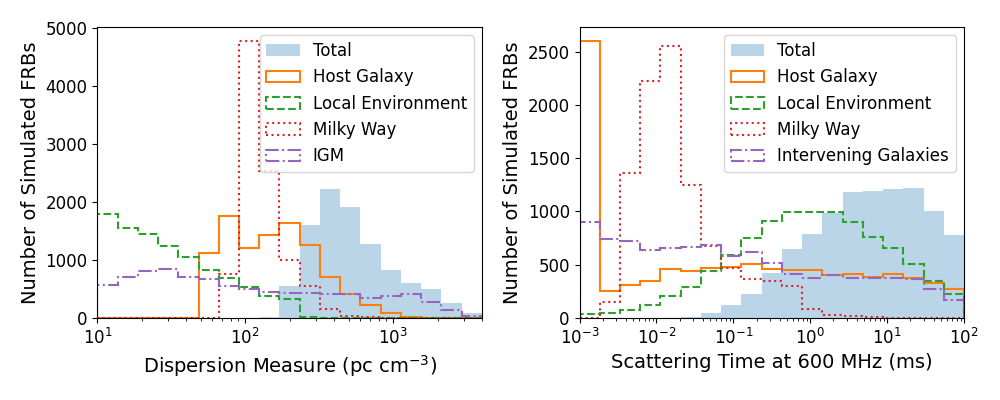}
    \caption{Simulated distributions of DMs and scattering times for the model found to be marginally consistent with Catalog 1 based on KS and AD tests. Histograms for the total DM and scattering time of the simulated FRBs are shown in blue. All other histograms show the DM and scattering contributions of different intervening media (see \S \ref{sec:prop}). For visualization purposes, scattering times $<10^{-3}$ ms are set to be equal to 10$^{-3}$ ms in the right panel. The contribution of the IGM is not shown in the right panel as we do not simulate scattering in the IGM (see \S \ref{sec:igm}). We remove FRBs with total scattering times $>100$ ms from the simulated distribution to allow for comparison with the selection-corrected scattering time distribution obtained for Catalog 1, which extends up to 100 ms.}
    \label{fig:contributions}
\end{figure}

The simulated DM and scattering distributions for the model we cannot exclude along with the corresponding distributions for Catalog 1 are shown in Figure \ref{fig:optimal} and \ref{fig:cdfoptimal}. The DM and scattering time distributions for the model are obtained by modeling different intervening media (see \S \ref{sec:prop}), the contributions of which are shown in Figure \ref{fig:contributions}. The conclusion of the aforementioned model being marginally consistent with the observations is subject to various assumptions that we test in \S \ref{sec:varyparameters}. We discuss the constraints we can place on the properties of the intervening media and on FRB progenitor models based on this result in \S \ref{sec:localconstraints}--\ref{sec:progenitors}.


\subsection{Varying Assumed Parameters}\label{sec:varyparameters}
The results reported in Table \ref{tab:models} are derived from simulations of intrinsic and observed properties of FRBs, modeling of which requires many assumptions. We investigate the robustness of our results to some of these assumptions by repeating our analysis for different values of the assumed parameters. 

The first set of assumptions is related to measured fluences in Catalog 1 being biased low as they are calculated assuming all bursts are detected at the most sensitive location along their transit. This systematic error propagates through to the assumed value of $E_\textrm{max}$ for the intrinsic energy distribution chosen based on the inferred energies of Catalog 1 events (as described in \S \ref{sec:energy}). Additionally, this one-sided systematic error also implies that the minimum detectable fluence assumed for the simulations (0.4 Jy ms; lowest-measured fluence in Catalog 1) could be an underestimate. 

We rerun the simulations for all population models twice. In the first iteration, we set $E_\textrm{max}$ to $10^{46}$ erg as compared to its previously assumed value of $10^{42}$ erg. While increasing E$_\mathrm{max}$ allows more events to be detectable at higher redshifts and hence higher DMs, we find that it does not significantly alter the overall DM distribution and the power-law index of the energy distribution. In the second iteration, we increase the minimum detectable fluence to twice the previously assumed value, setting it to be 0.8 Jy ms. The results for both these iterations confirm our initial conclusions. We find that a short GRB-like population in spiral galaxies with scattering arising in the local environment ($\mathrm{DM}_\textrm{Local} > 10$ pc cm$^{-3}$) and CGM of intervening galaxies ($f_\textrm{CGM} = 0.6$) still remains the only model even marginally consistent with Catalog 1.

The other assumed parameter that we vary is the DM contribution of the halo of the Milky Way and the host galaxy. While we initially assumed $\mathrm{DM}_\textrm{Halo} = 30$ pc cm$^{-3}$, we perform another iteration of the simulations by sampling halo DM uniformly in the range from 50 to 80 pc cm$^{-3}$ \citep{prochaska19}. We find the same model to be favored for this iteration as well. 

For all population models, the simulated scattering times are compared to catalog measurements which are derived assuming a power-law index of $-4$ for the frequency dependence. This choice is motivated by the measured scattering indices for pulsars and FRBs. While the mean value measured for 98 Galactic pulsars is $-3.9 \pm 0.2$ \citep{bhat04}, the measured indices for FRBs\footnote{\url{https://www.frbcat.org}} range from $-6$ to $-3.5$ \citep{petroff16}. All these measurements are consistent with an index of $-4$ due to the large associated uncertainties. However, these measurements could also be consistent with an index of $-4.4$, which is the theoretical expectation for a Kolmogorov medium with a negligible inner scale. 

While fitting the Catalog 1 bursts with a scattering timescale valid for a power-law index of $-4.4$ is outside the scope of this paper, we note that a difference of 0.4 in the index will modify the scattering time by a maximum of $(\nu_1/\nu_2)^{-0.4} \sim 30\%$. Here $\nu_1$ and $\nu_2$ are the lowest and highest observing frequencies, respectively. Since the PDF for the scattering time distribution in the catalog is evaluated for logarithmic bins, measured scattering times for an index of $-4.4$ would lie in the same or adjacent bin. The results of the comparison between the simulated distributions and Catalog 1 would thus not change significantly if a power-law index of $-4.4$ is assumed.

The burst-fitting process for Catalog 1 (see \S \ref{sec:observations}) also involves the assumption of an exponential scattering tail. An exponential tail is expected only if the scattering medium has a single characteristic scale of inhomogeneities \citep{ostashov77} and the scatter-broadened image of the source has a Gaussian brightness distribution \citep{cronyn70}. Deviations from exponential decay can arise if the electron density fluctuations have a power-law spectrum or the inner scale of the fluctuations is comparable to the diffractive scale \citep{rickett09,ostashov77}. Although we fit a single functional form to all bursts in order to simplify the fitting process, adopting a different functional form is unlikely to change our results. This is because minor variations in the inferred scattering timescales are not expected to significantly change the resulting PDF which is evaluated for logarithmic bins.

\subsection{Circumburst  Environment}\label{sec:localconstraints}
\begin{figure}[h!]
\centering
    \includegraphics[scale=0.6]{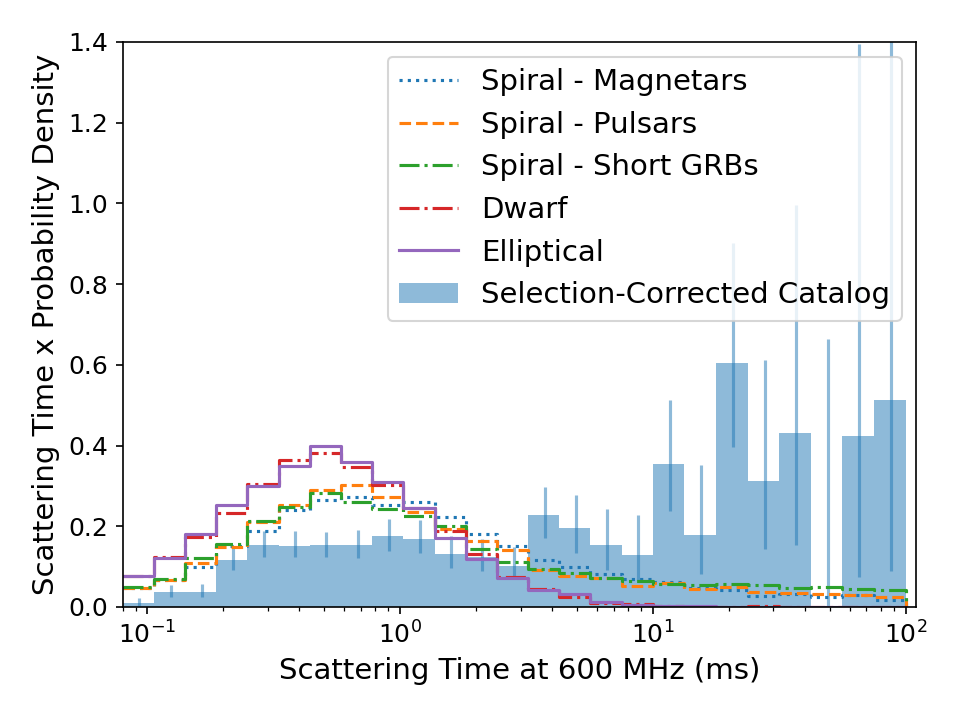}
    \caption{Scattering time distribution for population models in which the ISM of the host galaxy and the Milky Way are the dominant sources of scattering. The models assume FRBs to be located in different types of host galaxies. For spiral galaxies, three spatial distributions within the host are simulated, which emulate the magnetar, pulsar and short GRB populations. These models do not include DM or scattering contribution from the circumburst environment. The histograms represent the probability density function re-parameterized in terms of the natural logarithm of the scattering timescale. The scattering time distribution for the catalog is shown in blue. The catalog data are corrected for selection effects with the error bars showing the 68\% Poissonian confidence interval for each bin value. The simulation histograms show the average PDF over 100 Monte Carlo realizations of each model.}
    \label{fig:hostism}
\end{figure}

We find that none of the population models in which scattering originates only in the ISM of the Milky Way and the host galaxy can reproduce the observed scattering time distribution (see Table \ref{tab:models}). These models fail to match the large population of events with scattering times $>10$ ms, as shown in Figure \ref{fig:hostism}. This result confirms the conclusions of a population synthesis analysis conducted for the first 13 CHIME-detected bursts, which suggested that FRBs must have local environments with stronger scattering properties than the Milky Way ISM if all FRBs are located in spiral galaxies \citep{chime19a}. Similar to this work, the aforementioned analysis assumed that all FRBs are located in spiral galaxies, the ISM for which was simulated using the thin and thick disk, spiral arms and galactic center components of the NE2001 model. Additionally, our result is also consistent with the observation of two FRBs located in the outskirts of their host galaxies showing significant scattering \citep{day20}. 

\begin{figure}[h!]
\centering
    \includegraphics[scale=0.65]{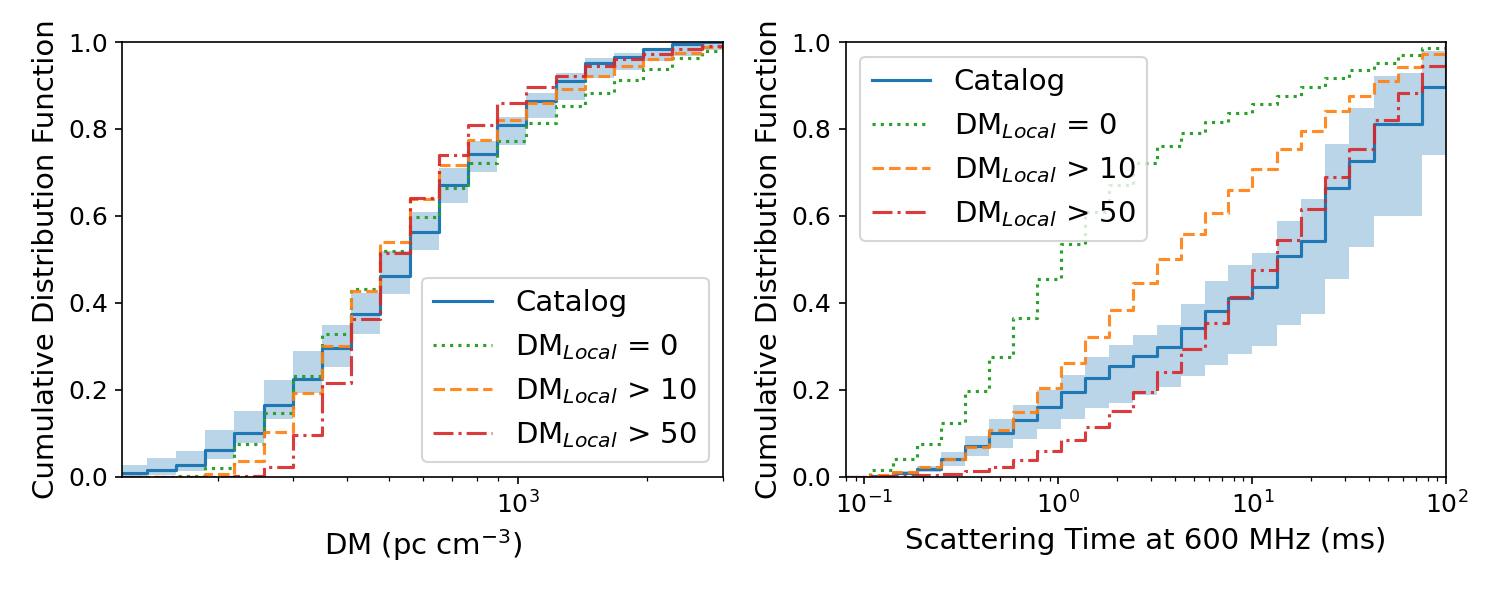}
    \caption{Cumulative distribution function for simulated DMs and scattering times for a short GRB-like spatial distribution of FRBs in spiral galaxies. Simulated distributions for three population models are shown here with the models differing in the contribution of the local environment to burst DMs and scattering times. The DMs of the local environment are drawn from a lognormal distribution (see \S \ref{sec:local}) with a minimum threshold as indicated in the legend. Increasing the local DM threshold results in higher scattering times, which reduces the discrepancy between the simulated and observed scattering time distribution but exacerbates the differences between the corresponding DM distributions at the low-DM end. The CDF for the simulated distributions are averaged over 100 Monte Carlo realizations for each model. The selection-corrected DM and scattering distributions in the catalog are plotted in blue. Uncertainties on the catalog CDF are determined in the same manner as for Figure \ref{fig:cdfoptimal}.}
    \label{fig:varylocal}
\end{figure}

We also test models in which FRBs inhabit dense local environments, the DMs and scattering times for which are simulated based on empirically derived values for the lines of sight of $\sim$100 Galactic pulsars. The additional scattering contributed by these environments reduces the discrepancy between the simulated and observed population as is indicated by higher p-values for the KS test (see columns for which $\mathrm{DM}_\textrm{Local} > 0.1$ pc cm$^{-3}$ in Table \ref{tab:models}). We also allow for circumburst environments with more extreme properties by introducing a higher threshold for the local DM contribution,  $\mathrm{DM}_\textrm{Local} > 10$ pc cm$^{-3}$. Although these models, on average, have higher scattering times and higher p-values for the KS test, they are still unable to explain the highly scattered bursts in the observed population, as shown in Figure \ref{fig:varylocal}.  

While further increasing the local DM threshold increases the total scattering timescales, it causes the overall DM distribution to be inconsistent with the catalog, as shown in Figure \ref{fig:varylocal} for a model in which $\mathrm{DM}_\textrm{Local} > 50$ pc cm$^{-3}$. Our results therefore suggest that the circumburst media must contribute more scattering per unit DM than typical Galactic plane environments such as \ion{H}{2} regions or star-formation complexes. This agrees with the results of a population study of scattering in Parkes FRBs which concluded that FRBs must inhabit a denser and more turbulent environment than a SGR 1935+2154-like magnetar \citep{hackstein20}. One caveat, however, is that scattering could be contributed by sources other than the local environment and the ISM of the Milky Way and the host galaxy. We discuss the CGM of intervening galaxies as one such source in \S \ref{sec:cgmconstraints}.

\subsection{Intervening Galaxies}\label{sec:cgmconstraints}
Introducing the CGM of intervening galaxies as an additional source of scattering reduces the discrepancy between the observed and simulated distributions. This is illustrated in Table \ref{tab:models}, where population models with no scattering arising in intervening galaxies show higher deviation from the observed distribution as compared to models in which this source of scattering is included. Among the population models we test, the only one we cannot rule out has scattering originating both in the circumburst environment and in intervening galaxies. This model is statistically preferred over one with scattering originating only in the circumburst environment because of two reasons. Firstly, the geometric weighting factor in Equation \ref{eq:sm}, $(s/D_A)(1-s/D_A)$, implies that material located midway between the source and the observer contributes most significantly to the observed scattering. Therefore, material of the same scattering strength would cause more intense scattering if it were located in an intervening galaxy instead of in the circumburst medium. Secondly, the sub-parsec scale clumps in the CGM of intervening galaxies \citep{vedantham19} have additional scattering strength, i.e., contribute more scattering per unit DM as compared to the simulated circumburst environments.

However, it could be that the CGM of intervening galaxies is not as turbulent as assumed in the model put forth by \citet{vedantham19}. \citet{ocker21} place an upper limit on the fluctuation parameter, which is proportional to ($\tau/\mathrm{DM}^2$), for the CGM of the Milky Way using scattering measurements of two FRBs. The fluctuation parameter that they infer is orders of magnitude lower than that proposed for the CGM by \citet{vedantham19}, suggesting that halos of other galaxies might not significantly contribute to FRB scattering. Recent observations of two FRBs whose sightlines intercept halos of other galaxies also support this claim \citep{cho20, simha20}. The aforementioned studies prompt us to reconsider the conclusions of \S \ref{sec:localconstraints}. If the proposed model of CGM scattering is correct, then the halos of intervening galaxies along with typical galactic plane environments can explain the scattering properties of CHIME/FRB detected sources. However, if the CGM of intervening galaxies is not as turbulent as is suggested by \citet{vedantham19}, then more extreme circumburst environments are required.

\subsection{Host Galaxies}\label{sec:hostresults}
We initially tested population models in which all FRBs are located in host galaxies of the
same type, in order to reduce the computational cost of the simulations. For this simplistic
scenario, the only model that we cannot formally exclude given the data assumes that all FRBs are located in spiral galaxies. However, observed host galaxies for localized FRBs seem to be drawn from a more heterogeneous sample. While the first repeating FRB was localized to a dwarf galaxy \citep{tendulkar17}, since then several spiral and elliptical hosts have been identified with varying rates of star formation \citep{bhandari20}. More recently, \citet{heintz20} studied the properties of the localized FRB sample and rejected the hypothesis that the population originates exclusively in late-type galaxies with $>3\sigma$ confidence.

To test whether the observed scenario of a heterogeneous sample of host galaxies is consistent with the DM and scattering distributions in the catalog, we rerun our simulations for the model that initially reproduced our observations. However, for this iteration, we place half of the FRB population in spiral galaxies and half in elliptical galaxies. We keep all other parameters constant as we simulate a short GRB-like population evolving with the star-formation rate with scattering arising in the local environment and in intervening galaxies. We find that the population distributed like short GRBs with a mixed distribution of FRB host galaxies is also marginally consistent with the catalog, with the constraints on scattering being the same as those in \S \ref{sec:cgmconstraints}. The mean p-value for the KS test comparing 100 realizations of the scattering time distribution for this model with the catalog is 10$^{-4}$, implying that the model is as likely as the one in which all FRBs have spiral hosts. As the scattering in both the models is mainly contributed by the circumburst environment and the CGM of intervening galaxies, the significance level does not change even after reducing the scattering contribution of the host ISM for half of the simulated FRBs. 

Another property of FRB host galaxies that has been constrained by recent population studies is their DM contribution. The constraints range from 50 to 270 pc cm$^{-3}$ (see, e.g., \citealt{gardenier20,yang17}) and include the contribution of the host galaxy and the circumburst environment. For simulated models consistent with Catalog 1, we find that the median rest-frame host DM ranges from 150 to 200 pc cm$^{-3}$. Our findings are consistent with a recent study by \citet{james21} of a sample of FRBs detected with the Parkes (Murriyang) and ASKAP telescopes. The sample also includes seven FRBs with confirmed host galaxies, based on which they infer an average rest-frame host DM of $145^{+60}_{-65}$ pc cm$^{-3}$.

More recently, an angular cross-correlation analysis of CHIME/FRB sources with cosmological galaxy catalogs has provided evidence for a sub-population of FRBs at z$\sim$0.4 with host DMs of $\sim$400 pc cm$^{-3}$ \citep{rafiei21}. The aforementioned host DM is estimated in the observer's frame and translates to a rest-frame DM of $\sim$560 pc cm$^{-3}$. However, \citet{rafiei21} clarify that their results do not suggest that majority of FRBs have host DMs greater than this value, implying that the median host DM for our simulations could be consistent with their findings. 

\citet{rafiei21} suggest that the high host DMs for this sub-population of FRBs could be explained if they are located near the centers of large (10$^{14}$ M$_\sun$) halos. Although we simulate the DM contribution of the host halo in our analysis, we assume it to be similar to the Milky Way (ranging from 30 to 80 pc cm$^{-3}$). We do not consider the scattering contribution of the host halo as it is expected to be $<$1 ms for a Milky Way-like galaxy (see \S \ref{sec:MW}). Scattering timescales could be larger for halos which are more massive than that of the Milky Way, which has a mass of $\sim10^{12} \textrm{M}_\sun$ \citep{posti19}. We do not simulate a sub-population of FRBs located near these large halos due to the uncertain scattering time contribution but note that it could reduce the discrepancy between the simulated and observed scattering time distributions.

\subsection{Progenitor Models}\label{sec:progenitors}
Our simulations favor a population of FRBs offset from their galaxy centers,  modeled based on the observed offsets for the short GRB population. Other progenitor models with offsets similar to those observed for short GRBs could also be consistent with our observations. This result is consistent with observations of FRBs that have been localized with sub-arcsecond precision. \citet{mannings20} study the offset distribution for eight such FRBs and find that it is consistent both with a population of core-collapse supernovae and short GRBs (see Figure 4 of their paper). 

\begin{figure}[h!]
\centering
    \includegraphics[scale=0.65]{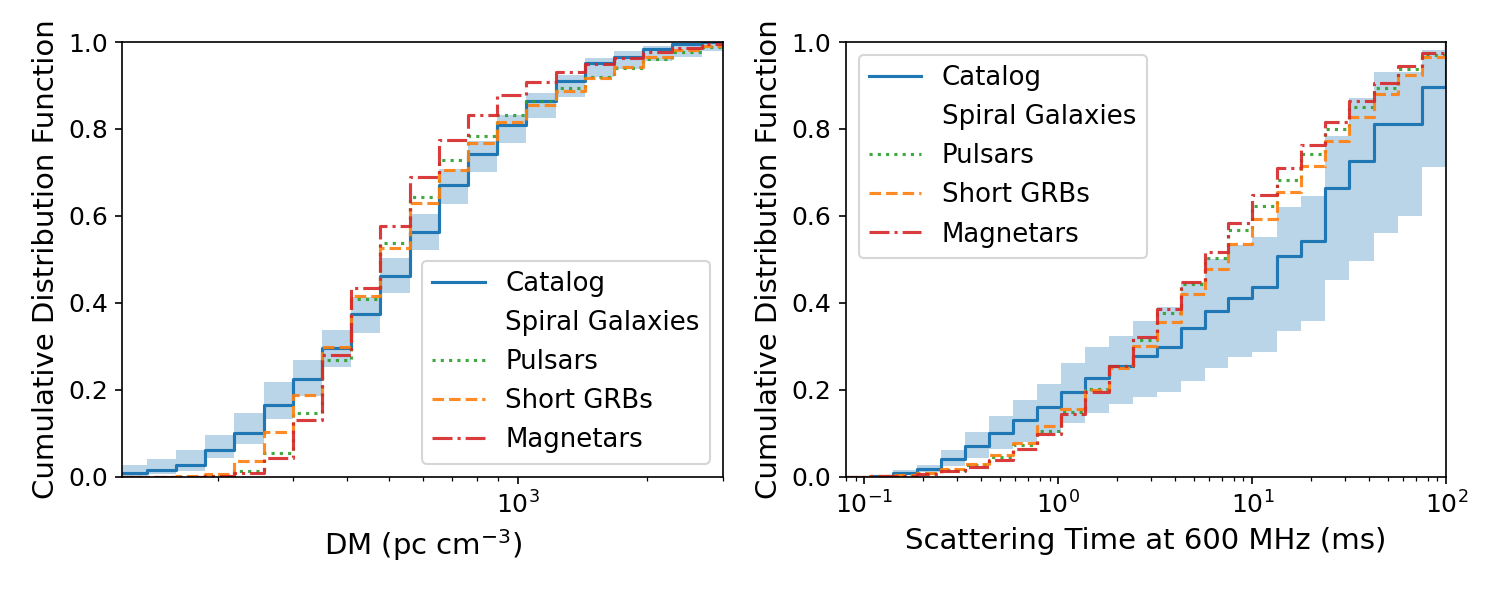}
    \caption{Cumulative distribution function of simulated DMs and scattering times for different potential FRB progenitors assumed to be located in spiral galaxies. Simulated population models assume that significant scattering is originating in the local environment with the local DM being drawn from a lognormal distribution (see \S \ref{sec:local}) having a minimum threshold of 10 pc cm$^{-3}$. Scattering is also assumed to be contributed by intervening galaxies with $f_\textrm{CGM} = 0.6$. The CDF for the simulated distributions are averaged over 100 Monte Carlo realizations for each model. The selection-corrected DM and scattering distributions in the catalog are plotted in blue. Uncertainties on the catalog CDF are determined in the same manner as for Figure \ref{fig:cdfoptimal}. While simulated scattering time distributions for some realizations of the pulsar and magnetar models are consistent with the corresponding catalog distribution, the simulated DM distributions for those realizations are found to be inconsistent with the catalog on the basis of a KS test.}
    \label{fig:progenitors}
\end{figure}

If the circumburst environment and intervening galaxies contribute significantly to scattering, the simulated scattering times for all tested spatial distributions are consistent with the catalog, as shown in Figure \ref{fig:progenitors}. However, spatial distributions resembling those of the pulsar and magnetar populations in the Milky Way can be ruled out due to a lack of low-DM FRBs (see Figure \ref{fig:progenitors}). The simulated DMs for a population modeled based on pulsars or magnetars are consistent with the catalog only if the contributions of the host galaxy, the Milky Way and the IGM are considered. The distributions become inconsistent if the DM contribution of dense circumburst environments is included, as shown in Figure \ref{fig:progenitors}. Since a population of FRBs offset from their galaxy centers has lower host DMs, on average, than a population distributed along the spiral arms, the simulated DM distribution for a short GRB-like population is statistically consistent with the catalog even with the inclusion of dense circumburst environments. 

It is important to note that we cannot rule out pulsars or magnetars as FRB progenitors if FRBs originate in environments with stronger scattering properties (such as a very young supernova remnant) than those simulated in our analysis. This is because higher scattering per unit DM in the local environment can potentially increase the number of low DM FRBs with high scattering times and reduce the discrepancy between the simulated and observed distributions in Figure \ref{fig:progenitors}. Although we test different spatial distributions in spiral galaxies, any spatial distribution of FRBs located in a mixed distribution of host galaxies could explain our observations, provided the FRBs inhabit circumburst environments with strong scattering properties. In this scenario, the elliptical and dwarf hosts might be able to explain the observed low DMs while the circumburst environments can reproduce the high scattering timescales.

\section{Summary and Conclusions}\label{sec:conclusion}
We have performed a population synthesis study to interpret the joint distribution of dispersion measures and scattering timescales for the first CHIME/FRB catalog. The study involved simulating FRB populations, specifically, their intrinsic properties and propagation effects arising in the Milky Way, the IGM, the CGM of intervening galaxies, the host galaxy and the circumburst environment. The simulated FRB populations were based on different models varying in their redshift distribution, host galaxy types and spatial distribution within the host. 

We compared the DM and scattering time distributions for the simulated FRB populations with the corresponding selection-corrected distributions for the catalog using KS and AD tests. For all population models that we simulate, we find that we cannot reproduce the observed scattering times if FRBs have circumburst environments with properties similar to those inferred for the local environments of Galactic pulsars. Based on this result, we infer that the circumburst media of FRBs must have more extreme properties than those of typical Galactic plane environments, thereby improving on the constraints that were derived based on the scattering times of the first 13 CHIME-detected FRBs \citep{chime19b}. 

We also test the possibility that sub-parsec scale cool ionized gas clumps in the circumgalactic medium of intervening galaxies could be contributing to FRB scattering \citep{vedantham19}. We find that we cannot rule out a model of FRBs for which scattering originates in both in the local environment and in intervening galaxies. If the proposed models for CGM scattering are correct, then this additional source of scattering relaxes the requirement of local environments of FRBs having more extreme properties than the ensemble of Galactic plane environments that we model.

While we cannot as yet determine the dominant host galaxy type using the observed
DM and scattering distributions, we place some constraints on FRB progenitor models. Our simulations favor a population of FRBs offset from their galaxy centers -- like the short GRB population -- over a population which is distributed along the spiral arms such as the magnetar and pulsar populations in the Milky Way. However, this result is dependent on the assumption that circumburst environments resemble those of Galactic pulsars. 

Another major caveat is that we simulate FRBs in different host galaxies, the electron density distributions of which are poorly understood. The assumed electron density models might not be able to adequately characterize the dispersive and scattering properties of these galaxies. Population studies of this kind therefore need to be augmented by analysis of scattering budgets of localized FRBs (see, e.g., \citealt{simha20,ocker21}). We encourage more such studies as the scattering time measurement for localized sources can be interpreted using additional information about the host galaxy, intervening galaxies and the local environment available from multi-wavelength follow-up observations. Such studies could then provide priors on the level of turbulence in other galaxies which in turn can enhance the robustness of population synthesis studies. 

In addition to causing temporal broadening, multi-path propagation can introduce other observable effects, namely, angular broadening and scintillation. Measurement of scintillation bandwidths and angular sizes of FRBs can be used to infer the location of the scattering material along the line of sight \citep{masui15,marcote17} and verify the conclusions presented in this paper. While angular broadening is only observable using Very Long Baseline Interferometry (VLBI), scintillation bandwidths have been measured for $\sim$10 bright FRBs detected with CHIME \citep{schoen21}. A population study of scintillation bandwidths for the CHIME/FRB sources is reserved for future work. 

It is also important to note that the aforementioned conclusions are derived for the full FRB population as the selection-corrected distributions for the CHIME/FRB catalog include both repeating and so-far non-repeating sources \citep{chime21}. The CHIME/FRB system is expected to detect more repeating FRB sources in the coming years which can allow for determination of the selection-corrected distributions exclusively for the repeating FRB population. Analyses similar to the one we report on here can then help discern whether repeating FRBs constitute a different population as compared to non-repeating sources based on their observed scattering properties. 

\acknowledgments
We acknowledge that CHIME is located on the traditional, ancestral, and unceded territory of the Syilx/Okanagan people.

We thank the anonymous referee for
comments that have improved the quality of this manuscript.

We thank the Dominion Radio Astrophysical Observatory, operated by the National Research Council Canada, for gracious hospitality and expertise. CHIME is funded by a grant from the Canada Foundation for Innovation (CFI) 2012 Leading Edge Fund (Project 31170) and by contributions from the provinces of British Columbia, Québec and Ontario. The CHIME/FRB Project, which enabled development in common with the CHIME/Pulsar instrument, is funded by a grant from the CFI 2015 Innovation Fund (Project 33213) and by contributions from the provinces of British Columbia and Québec, and by the Dunlap Institute for Astronomy and Astrophysics at the University of Toronto. Additional support was provided by the Canadian Institute for Advanced Research (CIFAR), McGill University and the McGill Space Institute thanks to the Trottier Family Foundation, and the University of British Columbia. The Dunlap Institute is funded through an endowment established by the David Dunlap family and the University of Toronto. Research at Perimeter Institute is supported by the Government of Canada through Industry Canada and by the Province of Ontario through the Ministry of Research \& Innovation. The National Radio Astronomy Observatory is a facility of the National Science Foundation (NSF) operated under cooperative agreement by Associated Universities, Inc. FRB research at UBC is supported by an NSERC Discovery Grant and by the Canadian Institute for Advanced Research. We thank Compute Canada, the McGill Center for High Performance Computing, and Calcul Qu\'ebec for provision and maintenance of the Beluga supercomputer and related resources.

P.C. is supported by an FRQNT Doctoral Research Award. V.M.K. holds the Lorne Trottier Chair in Astrophysics \& Cosmology and a Distinguished James McGill Professorship and receives support from an NSERC Discovery Grant and Herzberg Award, from an R. Howard Webster Foundation Fellowship from the Canadian Institute for Advanced Research (CIFAR), and from the FRQNT Centre de Recherche en Astrophysique du Quebec. S.M.R. is a CIFAR Fellow and is supported by the NSF Physics Frontiers Center award 1430284. M.B. is supported by an FRQNT Doctoral Research Award. B.M.G. is supported by an NSERC Discovery Grant (RGPIN-2015-05948), and by the Canada Research Chairs (CRC) program. C.L. was supported by the U.S. Department of Defense (DoD) through the National Defense Science \& Engineering Graduate Fellowship (NDSEG) Program. K.W.M. is supported by an NSF Grant (2008031). J.M.P is a Kavli Fellow. M.M. is supported by an NSERC PGS-D award. D.M. is a Banting Fellow. A.B.P is a McGill Space Institute (MSI) Fellow and a Fonds de Recherche du Quebec -- Nature et Technologies (FRQNT) postdoctoral fellow. E.P. acknowledges funding from an NWO Veni Fellowship. K.S. is supported by the NSF Graduate Research Fellowship Program.  

\vspace{5mm}
\facilities{CHIME}

%\software{PRESTO \citep{ransom2001}, \texttt{cdmt} \citep{2017A&C....18...40B}, 

\bibliography{sample63}{}

\begin{thebibliography}{}
\expandafter\ifx\csname natexlab\endcsname\relax\def\natexlab#1{#1}\fi
\providecommand{\url}[1]{\href{#1}{#1}}
\providecommand{\dodoi}[1]{doi:~\href{http://doi.org/#1}{\nolinkurl{#1}}}
\providecommand{\doeprint}[1]{\href{http://ascl.net/#1}{\nolinkurl{http://ascl.net/#1}}}
\providecommand{\doarXiv}[1]{\href{https://arxiv.org/abs/#1}{\nolinkurl{https://arxiv.org/abs/#1}}}

\bibitem[{{Bhandari} {et~al.}(2020){Bhandari}, {Sadler}, {Prochaska}, {Simha},
  {Ryder}, {Marnoch}, {Bannister}, {Macquart}, {Flynn}, {Shannon}, {Tejos},
  {Corro-Guerra}, {Day}, {Deller}, {Ekers}, {Lopez}, {Mahony}, {Nu{\~n}ez}, \&
  {Phillips}}]{bhandari20}
{Bhandari}, S., {Sadler}, E.~M., {Prochaska}, J.~X., {et~al.} 2020, \apjl, 895,
  L37, \dodoi{10.3847/2041-8213/ab672e}

\bibitem[{{Bhat} {et~al.}(2004){Bhat}, {Cordes}, {Camilo}, {Nice}, \&
  {Lorimer}}]{bhat04}
{Bhat}, N.~D.~R., {Cordes}, J.~M., {Camilo}, F., {Nice}, D.~J., \& {Lorimer},
  D.~R. 2004, \apj, 605, 759, \dodoi{10.1086/382680}

\bibitem[{{Blandford} \& {Narayan}(1985)}]{blandford85}
{Blandford}, R., \& {Narayan}, R. 1985, \mnras, 213, 591,
  \dodoi{10.1093/mnras/213.3.591}

\bibitem[{{Bochenek} {et~al.}(2020){Bochenek}, {Ravi}, {Belov}, {Hallinan},
  {Kocz}, {Kulkarni}, \& {McKenna}}]{bochenek20}
{Bochenek}, C.~D., {Ravi}, V., {Belov}, K.~V., {et~al.} 2020, \nat, 587, 59,
  \dodoi{10.1038/s41586-020-2872-x}

\bibitem[{{Caleb} {et~al.}(2016){Caleb}, {Flynn}, {Bailes}, {Barr}, {Hunstead},
  {Keane}, {Ravi}, \& {van Straten}}]{caleb16}
{Caleb}, M., {Flynn}, C., {Bailes}, M., {et~al.} 2016, \mnras, 458, 708,
  \dodoi{10.1093/mnras/stw175}

\bibitem[{{CHIME/FRB Collaboration} {et~al.}(2018){CHIME/FRB Collaboration},
  {Amiri}, {Bandura}, {Berger}, {Bhardwaj}, {Boyce}, {Boyle}, {Brar},
  {Burhanpurkar}, {Chawla}, {Chowdhury}, {Cliche}, {Cranmer}, {Cubranic},
  {Deng}, {Denman}, {Dobbs}, {Fandino}, {Fonseca}, {Gaensler}, {Giri},
  {Gilbert}, {Good}, {Guliani}, {Halpern}, {Hinshaw}, {H{\"o}fer}, {Josephy},
  {Kaspi}, {Landecker}, {Lang}, {Liao}, {Masui}, {Mena-Parra}, {Naidu},
  {Newburgh}, {Ng}, {Patel}, {Pen}, {Pinsonneault-Marotte}, {Pleunis}, {Rafiei
  Ravandi}, {Ransom}, {Renard}, {Scholz}, {Sigurdson}, {Siegel}, {Smith},
  {Stairs}, {Tendulkar}, {Vanderlinde}, \& {Wiebe}}]{chime18}
{CHIME/FRB Collaboration}, {Amiri}, M., {Bandura}, K., {et~al.} 2018, \apj,
  863, 48, \dodoi{10.3847/1538-4357/aad188}

\bibitem[{{CHIME/FRB Collaboration} {et~al.}(2019{\natexlab{a}}){CHIME/FRB
  Collaboration}, {Andersen}, {Bandura}, {Bhardwaj}, {Boubel}, {Boyce},
  {Boyle}, {Brar}, {Cassanelli}, {Chawla}, {Cubranic}, {Deng}, {Dobbs},
  {Fandino}, {Fonseca}, {Gaensler}, {Gilbert}, {Giri}, {Good}, {Halpern},
  {Hill}, {Hinshaw}, {H{\"o}fer}, {Josephy}, {Kaspi}, {Kothes}, {Landecker},
  {Lang}, {Li}, {Lin}, {Masui}, {Mena-Parra}, {Merryfield}, {Mckinven},
  {Michilli}, {Milutinovic}, {Naidu}, {Newburgh}, {Ng}, {Patel}, {Pen},
  {Pinsonneault-Marotte}, {Pleunis}, {Rafiei-Ravandi}, {Rahman}, {Ransom},
  {Renard}, {Scholz}, {Siegel}, {Singh}, {Smith}, {Stairs}, {Tendulkar},
  {Tretyakov}, {Vanderlinde}, {Yadav}, \& {Zwaniga}}]{chime19b}
{CHIME/FRB Collaboration}, {Andersen}, B.~C., {Bandura}, K., {et~al.}
  2019{\natexlab{a}}, \apjl, 885, L24, \dodoi{10.3847/2041-8213/ab4a80}

\bibitem[{{CHIME/FRB Collaboration} {et~al.}(2019{\natexlab{b}}){CHIME/FRB
  Collaboration}, {Amiri}, {Bandura}, {Bhardwaj}, {Boubel}, {Boyce}, {Boyle},
  {Brar}, {Burhanpurkar}, {Chawla}, {Cliche}, {Cubranic}, {Deng}, {Denman},
  {Dobbs}, {Fandino}, {Fonseca}, {Gaensler}, {Gilbert}, {Giri}, {Good},
  {Halpern}, {Hanna}, {Hill}, {Hinshaw}, {H{\"o}fer}, {Josephy}, {Kaspi},
  {Landecker}, {Lang}, {Masui}, {Mckinven}, {Mena-Parra}, {Merryfield},
  {Milutinovic}, {Moatti}, {Naidu}, {Newburgh}, {Ng}, {Patel}, {Pen},
  {Pinsonneault-Marotte}, {Pleunis}, {Rafiei-Ravandi}, {Ransom}, {Renard},
  {Scholz}, {Shaw}, {Siegel}, {Smith}, {Stairs}, {Tendulkar}, {Tretyakov},
  {Vanderlinde}, \& {Yadav}}]{chime19a}
{CHIME/FRB Collaboration}, {Amiri}, M., {Bandura}, K., {et~al.}
  2019{\natexlab{b}}, \nat, 566, 230, \dodoi{10.1038/s41586-018-0867-7}

\bibitem[{{CHIME/FRB Collaboration} {et~al.}(2020){CHIME/FRB Collaboration},
  {Andersen}, {Bandura}, {Bhardwaj}, {Bij}, {Boyce}, {Boyle}, {Brar},
  {Cassanelli}, {Chawla}, {Chen}, {Cliche}, {Cook}, {Cubranic}, {Curtin},
  {Denman}, {Dobbs}, {Dong}, {Fandino}, {Fonseca}, {Gaensler}, {Giri}, {Good},
  {Halpern}, {Hill}, {Hinshaw}, {H{\"o}fer}, {Josephy}, {Kania}, {Kaspi},
  {Landecker}, {Leung}, {Li}, {Lin}, {Masui}, {McKinven}, {Mena-Parra},
  {Merryfield}, {Meyers}, {Michilli}, {Milutinovic}, {Mirhosseini},
  {M{\"u}nchmeyer}, {Naidu}, {Newburgh}, {Ng}, {Patel}, {Pen},
  {Pinsonneault-Marotte}, {Pleunis}, {Quine}, {Rafiei-Ravandi}, {Rahman},
  {Ransom}, {Renard}, {Sanghavi}, {Scholz}, {Shaw}, {Shin}, {Siegel}, {Singh},
  {Smegal}, {Smith}, {Stairs}, {Tan}, {Tendulkar}, {Tretyakov}, {Vanderlinde},
  {Wang}, {Wulf}, \& {Zwaniga}}]{chime20}
{CHIME/FRB Collaboration}, {Andersen}, B.~C., {Bandura}, K.~M., {et~al.} 2020,
  \nat, 587, 54, \dodoi{10.1038/s41586-020-2863-y}

\bibitem[{{CHIME/FRB Collaboration} {et~al.}(2021){CHIME/FRB Collaboration},
  {:}, {Amiri}, {Andersen}, {Bandura}, {Berger}, {Bhardwaj}, {Boyce}, {Boyle},
  {Brar}, {Breitman}, {Cassanelli}, {Chawla}, {Chen}, {Cliche}, {Cook},
  {Cubranic}, {Curtin}, {Deng}, {Dobbs}, {Fengqiu}, {Dong}, {Eadie}, {Fandino},
  {Fonseca}, {Gaensler}, {Giri}, {Good}, {Halpern}, {Hill}, {Hinshaw},
  {Josephy}, {Kaczmarek}, {Kader}, {Kania}, {Kaspi}, {Landecker}, {Lang},
  {Leung}, {Li}, {Lin}, {Masui}, {Mckinven}, {Mena-Parra}, {Merryfield},
  {Meyers}, {Michilli}, {Milutinovic}, {Mirhosseini}, {M{\"u}nchmeyer},
  {Naidu}, {Newburgh}, {Ng}, {Patel}, {Pen}, {Petroff}, {Pinsonneault-Marotte},
  {Pleunis}, {Rafiei-Ravandi}, {Rahman}, {Ransom}, {Renard}, {Sanghavi},
  {Scholz}, {Shaw}, {Shin}, {Siegel}, {Sikora}, {Singh}, {Smith}, {Stairs},
  {Tan}, {Tendulkar}, {Vanderlinde}, {Wang}, {Wulf}, \& {Zwaniga}}]{chime21}
{CHIME/FRB Collaboration}, {:}, {Amiri}, M., {et~al.} 2021, arXiv e-prints,
  arXiv:2106.04352.
\newblock \doarXiv{2106.04352}

\bibitem[{{Chittidi} {et~al.}(2020){Chittidi}, {Simha}, {Mannings},
  {Prochaska}, {Rafelski}, {Neeleman}, {Macquart}, {Tejos}, {Jorgenson},
  {Ryder}, {Day}, {Marnoch}, {Bhandari}, {Deller}, {Qiu}, {Bannister},
  {Shannon}, \& {Heintz}}]{chittidi20}
{Chittidi}, J.~S., {Simha}, S., {Mannings}, A., {et~al.} 2020, arXiv e-prints,
  arXiv:2005.13158.
\newblock \doarXiv{2005.13158}

\bibitem[{{Cho} {et~al.}(2020){Cho}, {Macquart}, {Shannon}, {Deller},
  {Morrison}, {Ekers}, {Bannister}, {Farah}, {Qiu}, {Sammons}, {Bailes},
  {Bhandari}, {Day}, {James}, {Phillips}, {Prochaska}, \& {Tuthill}}]{cho20}
{Cho}, H., {Macquart}, J.-P., {Shannon}, R.~M., {et~al.} 2020, \apjl, 891, L38,
  \dodoi{10.3847/2041-8213/ab7824}

\bibitem[{{Connor} {et~al.}(2016){Connor}, {Sievers}, \& {Pen}}]{connor16}
{Connor}, L., {Sievers}, J., \& {Pen}, U.-L. 2016, \mnras, 458, L19,
  \dodoi{10.1093/mnrasl/slv124}

\bibitem[{{Cordes} \& {Chatterjee}(2019)}]{cordes19}
{Cordes}, J.~M., \& {Chatterjee}, S. 2019, \araa, 57, 417,
  \dodoi{10.1146/annurev-astro-091918-104501}

\bibitem[{{Cordes} \& {Lazio}(2002)}]{cordes02}
{Cordes}, J.~M., \& {Lazio}, T.~J.~W. 2002, arXiv e-prints, astro.
\newblock \doarXiv{astro-ph/0207156}

\bibitem[{{Cordes} \& {Lazio}(2003)}]{cordes03}
---. 2003, arXiv e-prints, astro.
\newblock \doarXiv{astro-ph/0301598}

\bibitem[{{Cordes} \& {Wasserman}(2016)}]{cordes16b}
{Cordes}, J.~M., \& {Wasserman}, I. 2016, \mnras, 457, 232,
  \dodoi{10.1093/mnras/stv2948}

\bibitem[{{Cordes} {et~al.}(2016){Cordes}, {Wharton}, {Spitler}, {Chatterjee},
  \& {Wasserman}}]{cordes16}
{Cordes}, J.~M., {Wharton}, R.~S., {Spitler}, L.~G., {Chatterjee}, S., \&
  {Wasserman}, I. 2016, arXiv e-prints, arXiv:1605.05890.
\newblock \doarXiv{1605.05890}

\bibitem[{{Cronyn}(1970)}]{cronyn70}
{Cronyn}, W.~M. 1970, Science, 168, 1453, \dodoi{10.1126/science.168.3938.1453}

\bibitem[{{Day} {et~al.}(2020){Day}, {Deller}, {Shannon}, {Qiu}, {Bannister},
  {Bhandari}, {Ekers}, {Flynn}, {James}, {Macquart}, {Mahony}, {Phillips}, \&
  {Xavier Prochaska}}]{day20}
{Day}, C.~K., {Deller}, A.~T., {Shannon}, R.~M., {et~al.} 2020, \mnras, 497,
  3335, \dodoi{10.1093/mnras/staa2138}

\bibitem[{{Dolag} {et~al.}(2015){Dolag}, {Gaensler}, {Beck}, \&
  {Beck}}]{dolag15}
{Dolag}, K., {Gaensler}, B.~M., {Beck}, A.~M., \& {Beck}, M.~C. 2015, \mnras,
  451, 4277, \dodoi{10.1093/mnras/stv1190}

\bibitem[{{Faucher-Gigu{\`e}re} \& {Kaspi}(2006)}]{fgk06}
{Faucher-Gigu{\`e}re}, C.-A., \& {Kaspi}, V.~M. 2006, \apj, 643, 332,
  \dodoi{10.1086/501516}

\bibitem[{Flury(1990)}]{flury90}
Flury, B.~D. 1990, SIAM Review, 32, 474, \dodoi{10.1137/1032082}

\bibitem[{{Fong} \& {Berger}(2013)}]{fong13}
{Fong}, W., \& {Berger}, E. 2013, \apj, 776, 18,
  \dodoi{10.1088/0004-637X/776/1/18}

\bibitem[{{Fonseca} {et~al.}(2020){Fonseca}, {Andersen}, {Bhardwaj}, {Chawla},
  {Good}, {Josephy}, {Kaspi}, {Masui}, {Mckinven}, {Michilli}, {Pleunis},
  {Shin}, {Tendulkar}, {Bandura}, {Boyle}, {Brar}, {Cassanelli}, {Cubranic},
  {Dobbs}, {Dong}, {Gaensler}, {Hinshaw}, {Landecker}, {Leung}, {Li}, {Lin},
  {Mena-Parra}, {Merryfield}, {Naidu}, {Ng}, {Patel}, {Pen}, {Rafiei-Ravandi},
  {Rahman}, {Ransom}, {Scholz}, {Smith}, {Stairs}, {Vanderlinde}, {Yadav}, \&
  {Zwaniga}}]{fonseca20}
{Fonseca}, E., {Andersen}, B.~C., {Bhardwaj}, M., {et~al.} 2020, \apjl, 891,
  L6, \dodoi{10.3847/2041-8213/ab7208}

\bibitem[{{Gardenier} \& {van Leeuwen}(2020)}]{gardenier20}
{Gardenier}, D.~W., \& {van Leeuwen}, J. 2020, arXiv e-prints,
  arXiv:2012.06396.
\newblock \doarXiv{2012.06396}

\bibitem[{{Gardenier} {et~al.}(2019){Gardenier}, {van Leeuwen}, {Connor}, \&
  {Petroff}}]{gardenier19}
{Gardenier}, D.~W., {van Leeuwen}, J., {Connor}, L., \& {Petroff}, E. 2019,
  \aap, 632, A125, \dodoi{10.1051/0004-6361/201936404}

\bibitem[{{Hackstein} {et~al.}(2020){Hackstein}, {Br{\"u}ggen}, {Vazza}, \&
  {Rodrigues}}]{hackstein20}
{Hackstein}, S., {Br{\"u}ggen}, M., {Vazza}, F., \& {Rodrigues}, L.~F.~S. 2020,
  \mnras, 498, 4811, \dodoi{10.1093/mnras/staa2572}

\bibitem[{{Heintz} {et~al.}(2020){Heintz}, {Prochaska}, {Simha}, {Platts},
  {Fong}, {Tejos}, {Ryder}, {Aggerwal}, {Bhandari}, {Day}, {Deller},
  {Kilpatrick}, {Law}, {Macquart}, {Mannings}, {Marnoch}, {Sadler}, \&
  {Shannon}}]{heintz20}
{Heintz}, K.~E., {Prochaska}, J.~X., {Simha}, S., {et~al.} 2020, \apj, 903,
  152, \dodoi{10.3847/1538-4357/abb6fb}

\bibitem[{{Ioka}(2003)}]{ioka03}
{Ioka}, K. 2003, \apjl, 598, L79, \dodoi{10.1086/380598}

\bibitem[{{Isobe} {et~al.}(1990){Isobe}, {Feigelson}, {Akritas}, \&
  {Babu}}]{isobe90}
{Isobe}, T., {Feigelson}, E.~D., {Akritas}, M.~G., \& {Babu}, G.~J. 1990, \apj,
  364, 104, \dodoi{10.1086/169390}

\bibitem[{{James} {et~al.}(2021){James}, {Prochaska}, {Macquart},
  {North-Hickey}, {Bannister}, \& {Dunning}}]{james21}
{James}, C.~W., {Prochaska}, J.~X., {Macquart}, J.~P., {et~al.} 2021, arXiv
  e-prints, arXiv:2101.07998.
\newblock \doarXiv{2101.07998}

\bibitem[{{James} {et~al.}(2004){James}, {Shane}, {Beckman}, {Cardwell},
  {Collins}, {Etherton}, {de Jong}, {Fathi}, {Knapen}, {Peletier}, {Percival},
  {Pollacco}, {Seigar}, {Stedman}, \& {Steele}}]{james04}
{James}, P.~A., {Shane}, N.~S., {Beckman}, J.~E., {et~al.} 2004, \aap, 414, 23,
  \dodoi{10.1051/0004-6361:20031568}

\bibitem[{{Johnston} \& {Romani}(2003)}]{johnston03}
{Johnston}, S., \& {Romani}, R.~W. 2003, \apjl, 590, L95,
  \dodoi{10.1086/376826}

\bibitem[{{Josephy} {et~al.}(2019){Josephy}, {Chawla}, {Fonseca}, {Ng},
  {Patel}, {Pleunis}, {Scholz}, {Andersen}, {Bandura}, {Bhardwaj}, {Boyce},
  {Boyle}, {Brar}, {Cubranic}, {Dobbs}, {Gaensler}, {Gill}, {Giri}, {Good},
  {Halpern}, {Hinshaw}, {Kaspi}, {Landecker}, {Lang}, {Lin}, {Masui},
  {Mckinven}, {Mena-Parra}, {Merryfield}, {Michilli}, {Milutinovic}, {Naidu},
  {Pen}, {Rafiei-Ravandi}, {Rahman}, {Ransom}, {Renard}, {Siegel}, {Smith},
  {Stairs}, {Tendulkar}, {Vanderlinde}, {Yadav}, \& {Zwaniga}}]{josephy19}
{Josephy}, A., {Chawla}, P., {Fonseca}, E., {et~al.} 2019, \apjl, 882, L18,
  \dodoi{10.3847/2041-8213/ab2c00}

\bibitem[{{Katz}(2016)}]{katz16}
{Katz}, J.~I. 2016, \apj, 818, 19, \dodoi{10.3847/0004-637X/818/1/19}

\bibitem[{{Keating} \& {Pen}(2020)}]{keating20}
{Keating}, L.~C., \& {Pen}, U.-L. 2020, \mnras, 496, L106,
  \dodoi{10.1093/mnrasl/slaa095}

\bibitem[{{Kiel} {et~al.}(2010){Kiel}, {Hurley}, \& {Bailes}}]{kiel10}
{Kiel}, P.~D., {Hurley}, J.~R., \& {Bailes}, M. 2010, \mnras, 406, 656,
  \dodoi{10.1111/j.1365-2966.2010.16717.x}

\bibitem[{{Lorimer} {et~al.}(2006){Lorimer}, {Faulkner}, {Lyne}, {Manchester},
  {Kramer}, {McLaughlin}, {Hobbs}, {Possenti}, {Stairs}, {Camilo}, {Burgay},
  {D'Amico}, {Corongiu}, \& {Crawford}}]{lorimer06}
{Lorimer}, D.~R., {Faulkner}, A.~J., {Lyne}, A.~G., {et~al.} 2006, \mnras, 372,
  777, \dodoi{10.1111/j.1365-2966.2006.10887.x}

\bibitem[{{Macquart} \& {Ekers}(2018)}]{macquart18}
{Macquart}, J.~P., \& {Ekers}, R. 2018, \mnras, 480, 4211,
  \dodoi{10.1093/mnras/sty2083}

\bibitem[{{Macquart} \& {Koay}(2013)}]{macquart13}
{Macquart}, J.-P., \& {Koay}, J.~Y. 2013, \apj, 776, 125,
  \dodoi{10.1088/0004-637X/776/2/125}

\bibitem[{{Macquart} {et~al.}(2020){Macquart}, {Prochaska}, {McQuinn},
  {Bannister}, {Bhandari}, {Day}, {Deller}, {Ekers}, {James}, {Marnoch},
  {Os{\l}owski}, {Phillips}, {Ryder}, {Scott}, {Shannon}, \&
  {Tejos}}]{macquart20}
{Macquart}, J.~P., {Prochaska}, J.~X., {McQuinn}, M., {et~al.} 2020, \nat, 581,
  391, \dodoi{10.1038/s41586-020-2300-2}

\bibitem[{{Madau} \& {Dickinson}(2014)}]{madau14}
{Madau}, P., \& {Dickinson}, M. 2014, \araa, 52, 415,
  \dodoi{10.1146/annurev-astro-081811-125615}

\bibitem[{{Mannings} {et~al.}(2020){Mannings}, {Fong}, {Simha}, {Prochaska},
  {Rafelski}, {Kilpatrick}, {Tejos}, {Heintz}, {Bhandari}, {Day}, {Deller},
  {Ryder}, {Shannon}, \& {Tendulkar}}]{mannings20}
{Mannings}, A.~G., {Fong}, W.-f., {Simha}, S., {et~al.} 2020, arXiv e-prints,
  arXiv:2012.11617.
\newblock \doarXiv{2012.11617}

\bibitem[{{Marcote} {et~al.}(2017){Marcote}, {Paragi}, {Hessels}, {Keimpema},
  {van Langevelde}, {Huang}, {Bassa}, {Bogdanov}, {Bower}, {Burke-Spolaor},
  {Butler}, {Campbell}, {Chatterjee}, {Cordes}, {Demorest}, {Garrett}, {Ghosh},
  {Kaspi}, {Law}, {Lazio}, {McLaughlin}, {Ransom}, {Salter}, {Scholz},
  {Seymour}, {Siemion}, {Spitler}, {Tendulkar}, \& {Wharton}}]{marcote17}
{Marcote}, B., {Paragi}, Z., {Hessels}, J.~W.~T., {et~al.} 2017, \apjl, 834,
  L8, \dodoi{10.3847/2041-8213/834/2/L8}

\bibitem[{{Marcote} {et~al.}(2020){Marcote}, {Nimmo}, {Hessels}, {Tendulkar},
  {Bassa}, {Paragi}, {Keimpema}, {Bhardwaj}, {Karuppusamy}, {Kaspi}, {Law},
  {Michilli}, {Aggarwal}, {Andersen}, {Archibald}, {Bandura}, {Bower}, {Boyle},
  {Brar}, {Burke-Spolaor}, {Butler}, {Cassanelli}, {Chawla}, {Demorest},
  {Dobbs}, {Fonseca}, {Giri}, {Good}, {Gourdji}, {Josephy}, {Kirichenko},
  {Kirsten}, {Landecker}, {Lang}, {Lazio}, {Li}, {Lin}, {Linford}, {Masui},
  {Mena-Parra}, {Naidu}, {Ng}, {Patel}, {Pen}, {Pleunis}, {Rafiei-Ravandi},
  {Rahman}, {Renard}, {Scholz}, {Siegel}, {Smith}, {Stairs}, {Vanderlinde}, \&
  {Zwaniga}}]{marcote20}
{Marcote}, B., {Nimmo}, K., {Hessels}, J.~W.~T., {et~al.} 2020, \nat, 577, 190,
  \dodoi{10.1038/s41586-019-1866-z}

\bibitem[{{Margalit} {et~al.}(2019){Margalit}, {Berger}, \&
  {Metzger}}]{margalit19}
{Margalit}, B., {Berger}, E., \& {Metzger}, B.~D. 2019, \apj, 886, 110,
  \dodoi{10.3847/1538-4357/ab4c31}

\bibitem[{{Margalit} \& {Metzger}(2018)}]{margalit18}
{Margalit}, B., \& {Metzger}, B.~D. 2018, \apjl, 868, L4,
  \dodoi{10.3847/2041-8213/aaedad}

\bibitem[{Massey~Jr.(1951)}]{massey51}
Massey~Jr., F.~J. 1951, Journal of the American Statistical Association, 46,
  68, \dodoi{10.1080/01621459.1951.10500769}

\bibitem[{{Masui} {et~al.}(2015){Masui}, {Lin}, {Sievers}, {Anderson}, {Chang},
  {Chen}, {Ganguly}, {Jarvis}, {Kuo}, {Li}, {Liao}, {McLaughlin}, {Pen},
  {Peterson}, {Roman}, {Timbie}, {Voytek}, \& {Yadav}}]{masui15}
{Masui}, K., {Lin}, H.-H., {Sievers}, J., {et~al.} 2015, \nat, 528, 523,
  \dodoi{10.1038/nature15769}

\bibitem[{{Masui} \& {Sigurdson}(2015)}]{masui15b}
{Masui}, K.~W., \& {Sigurdson}, K. 2015, \prl, 115, 121301,
  \dodoi{10.1103/PhysRevLett.115.121301}

\bibitem[{{McQuinn}(2014)}]{mcquinn14}
{McQuinn}, M. 2014, \apjl, 780, L33, \dodoi{10.1088/2041-8205/780/2/L33}

\bibitem[{{Metzger} {et~al.}(2019){Metzger}, {Margalit}, \&
  {Sironi}}]{metzger19}
{Metzger}, B.~D., {Margalit}, B., \& {Sironi}, L. 2019, \mnras, 485, 4091,
  \dodoi{10.1093/mnras/stz700}

\bibitem[{{Moustakas} {et~al.}(2013){Moustakas}, {Coil}, {Aird}, {Blanton},
  {Cool}, {Eisenstein}, {Mendez}, {Wong}, {Zhu}, \& {Arnouts}}]{moustakas13}
{Moustakas}, J., {Coil}, A.~L., {Aird}, J., {et~al.} 2013, \apj, 767, 50,
  \dodoi{10.1088/0004-637X/767/1/50}

\bibitem[{{Niu} {et~al.}(2021){Niu}, {Aggarwal}, {Li}, {Zhang}, {Chatterjee},
  {Tsai}, {Yu}, {Law}, {Burke-Spolaor}, {Cordes}, {Zhang}, {Ocker}, {Yao},
  {Wang}, {Feng}, {Niino}, {Bochenek}, {Cruces}, {Connor}, {Jiang}, {Dai},
  {Luo}, {Li}, {Miao}, {Niu}, {Anna-Thomas}, {Sydnor}, {Stern}, {Wang}, {Yuan},
  {Yue}, {Zhou}, {Yan}, {Zhu}, \& {Zhang}}]{niu21}
{Niu}, C.~H., {Aggarwal}, K., {Li}, D., {et~al.} 2021, arXiv e-prints,
  arXiv:2110.07418.
\newblock \doarXiv{2110.07418}

\bibitem[{{Ocker} {et~al.}(2020){Ocker}, {Cordes}, \& {Chatterjee}}]{ocker20}
{Ocker}, S.~K., {Cordes}, J.~M., \& {Chatterjee}, S. 2020, \apj, 897, 124,
  \dodoi{10.3847/1538-4357/ab98f9}

\bibitem[{{Ocker} {et~al.}(2021){Ocker}, {Cordes}, \& {Chatterjee}}]{ocker21}
---. 2021, \apj, 911, 102, \dodoi{10.3847/1538-4357/abeb6e}

\bibitem[{{Olausen} \& {Kaspi}(2014)}]{olausen14}
{Olausen}, S.~A., \& {Kaspi}, V.~M. 2014, \apjs, 212, 6,
  \dodoi{10.1088/0067-0049/212/1/6}

\bibitem[{{Ostashov} \& {Shishov}(1977)}]{ostashov77}
{Ostashov}, V.~E., \& {Shishov}, V.~I. 1977, Radiofizika, 20, 842

\bibitem[{{Petroff} {et~al.}(2019){Petroff}, {Hessels}, \&
  {Lorimer}}]{petroff19}
{Petroff}, E., {Hessels}, J.~W.~T., \& {Lorimer}, D.~R. 2019, \aapr, 27, 4,
  \dodoi{10.1007/s00159-019-0116-6}

\bibitem[{{Petroff} {et~al.}(2016){Petroff}, {Barr}, {Jameson}, {Keane},
  {Bailes}, {Kramer}, {Morello}, {Tabbara}, \& {van Straten}}]{petroff16}
{Petroff}, E., {Barr}, E.~D., {Jameson}, A., {et~al.} 2016, \pasa, 33, e045,
  \dodoi{10.1017/pasa.2016.35}

\bibitem[{{Piro}(2016)}]{piro16}
{Piro}, A.~L. 2016, \apjl, 824, L32, \dodoi{10.3847/2041-8205/824/2/L32}

\bibitem[{{Planck Collaboration} {et~al.}(2016){Planck Collaboration}, {Ade},
  {Aghanim}, {Arnaud}, {Ashdown}, {Aumont}, {Baccigalupi}, {Banday},
  {Barreiro}, {Bartlett}, {Bartolo}, {Battaner}, {Battye}, {Benabed},
  {Beno{\^\i}t}, {Benoit-L{\'e}vy}, {Bernard}, {Bersanelli}, {Bielewicz},
  {Bock}, {Bonaldi}, {Bonavera}, {Bond}, {Borrill}, {Bouchet}, {Boulanger},
  {Bucher}, {Burigana}, {Butler}, {Calabrese}, {Cardoso}, {Catalano},
  {Challinor}, {Chamballu}, {Chary}, {Chiang}, {Chluba}, {Christensen},
  {Church}, {Clements}, {Colombi}, {Colombo}, {Combet}, {Coulais}, {Crill},
  {Curto}, {Cuttaia}, {Danese}, {Davies}, {Davis}, {de Bernardis}, {de Rosa},
  {de Zotti}, {Delabrouille}, {D{\'e}sert}, {Di Valentino}, {Dickinson},
  {Diego}, {Dolag}, {Dole}, {Donzelli}, {Dor{\'e}}, {Douspis}, {Ducout},
  {Dunkley}, {Dupac}, {Efstathiou}, {Elsner}, {En{\ss}lin}, {Eriksen},
  {Farhang}, {Fergusson}, {Finelli}, {Forni}, {Frailis}, {Fraisse},
  {Franceschi}, {Frejsel}, {Galeotta}, {Galli}, {Ganga}, {Gauthier}, {Gerbino},
  {Ghosh}, {Giard}, {Giraud-H{\'e}raud}, {Giusarma}, {Gjerl{\o}w},
  {Gonz{\'a}lez-Nuevo}, {G{\'o}rski}, {Gratton}, {Gregorio}, {Gruppuso},
  {Gudmundsson}, {Hamann}, {Hansen}, {Hanson}, {Harrison}, {Helou},
  {Henrot-Versill{\'e}}, {Hern{\'a}ndez-Monteagudo}, {Herranz}, {Hildebrandt},
  {Hivon}, {Hobson}, {Holmes}, {Hornstrup}, {Hovest}, {Huang}, {Huffenberger},
  {Hurier}, {Jaffe}, {Jaffe}, {Jones}, {Juvela}, {Keih{\"a}nen}, {Keskitalo},
  {Kisner}, {Kneissl}, {Knoche}, {Knox}, {Kunz}, {Kurki-Suonio}, {Lagache},
  {L{\"a}hteenm{\"a}ki}, {Lamarre}, {Lasenby}, {Lattanzi}, {Lawrence}, {Leahy},
  {Leonardi}, {Lesgourgues}, {Levrier}, {Lewis}, {Liguori}, {Lilje},
  {Linden-V{\o}rnle}, {L{\'o}pez-Caniego}, {Lubin}, {Mac{\'\i}as-P{\'e}rez},
  {Maggio}, {Maino}, {Mandolesi}, {Mangilli}, {Marchini}, {Maris}, {Martin},
  {Martinelli}, {Mart{\'\i}nez-Gonz{\'a}lez}, {Masi}, {Matarrese}, {McGehee},
  {Meinhold}, {Melchiorri}, {Melin}, {Mendes}, {Mennella}, {Migliaccio},
  {Millea}, {Mitra}, {Miville-Desch{\^e}nes}, {Moneti}, {Montier}, {Morgante},
  {Mortlock}, {Moss}, {Munshi}, {Murphy}, {Naselsky}, {Nati}, {Natoli},
  {Netterfield}, {N{\o}rgaard-Nielsen}, {Noviello}, {Novikov}, {Novikov},
  {Oxborrow}, {Paci}, {Pagano}, {Pajot}, {Paladini}, {Paoletti}, {Partridge},
  {Pasian}, {Patanchon}, {Pearson}, {Perdereau}, {Perotto}, {Perrotta},
  {Pettorino}, {Piacentini}, {Piat}, {Pierpaoli}, {Pietrobon}, {Plaszczynski},
  {Pointecouteau}, {Polenta}, {Popa}, {Pratt}, {Pr{\'e}zeau}, {Prunet},
  {Puget}, {Rachen}, {Reach}, {Rebolo}, {Reinecke}, {Remazeilles}, {Renault},
  {Renzi}, {Ristorcelli}, {Rocha}, {Rosset}, {Rossetti}, {Roudier},
  {Rouill{\'e} d'Orfeuil}, {Rowan-Robinson}, {Rubi{\~n}o-Mart{\'\i}n},
  {Rusholme}, {Said}, {Salvatelli}, {Salvati}, {Sandri}, {Santos},
  {Savelainen}, {Savini}, {Scott}, {Seiffert}, {Serra}, {Shellard}, {Spencer},
  {Spinelli}, {Stolyarov}, {Stompor}, {Sudiwala}, {Sunyaev}, {Sutton},
  {Suur-Uski}, {Sygnet}, {Tauber}, {Terenzi}, {Toffolatti}, {Tomasi},
  {Tristram}, {Trombetti}, {Tucci}, {Tuovinen}, {T{\"u}rler}, {Umana},
  {Valenziano}, {Valiviita}, {Van Tent}, {Vielva}, {Villa}, {Wade}, {Wandelt},
  {Wehus}, {White}, {White}, {Wilkinson}, {Yvon}, {Zacchei}, \&
  {Zonca}}]{planck16}
{Planck Collaboration}, {Ade}, P.~A.~R., {Aghanim}, N., {et~al.} 2016, \aap,
  594, A13, \dodoi{10.1051/0004-6361/201525830}

\bibitem[{{Platts} {et~al.}(2020){Platts}, {Prochaska}, \& {Law}}]{platts20}
{Platts}, E., {Prochaska}, J.~X., \& {Law}, C.~J. 2020, \apjl, 895, L49,
  \dodoi{10.3847/2041-8213/ab930a}

\bibitem[{{Platts} {et~al.}(2019){Platts}, {Weltman}, {Walters}, {Tendulkar},
  {Gordin}, \& {Kandhai}}]{platts19}
{Platts}, E., {Weltman}, A., {Walters}, A., {et~al.} 2019, \physrep, 821, 1,
  \dodoi{10.1016/j.physrep.2019.06.003}

\bibitem[{{Pleunis} {et~al.}(2021){Pleunis}, {Good}, {Kaspi}, {Mckinven},
  {Ransom}, {Scholz}, {Bandura}, {Bhardwaj}, {Boyle}, {Brar}, {Cassanelli},
  {Chawla}, {Fengqiu}, {Dong}, {Fonseca}, {Gaensler}, {Josephy}, {Kaczmarek},
  {Leung}, {Lin}, {Masui}, {Mena-Parra}, {Michilli}, {Ng}, {Patel},
  {Rafiei-Ravandi}, {Rahman}, {Sanghavi}, {Shin}, {Smith}, {Stairs}, \&
  {Tendulkar}}]{pleunis21}
{Pleunis}, Z., {Good}, D.~C., {Kaspi}, V.~M., {et~al.} 2021, arXiv e-prints,
  arXiv:2106.04356.
\newblock \doarXiv{2106.04356}

\bibitem[{{Posti} \& {Helmi}(2019)}]{posti19}
{Posti}, L., \& {Helmi}, A. 2019, \aap, 621, A56,
  \dodoi{10.1051/0004-6361/201833355}

\bibitem[{{Prochaska} \& {Zheng}(2019)}]{prochaska19}
{Prochaska}, J.~X., \& {Zheng}, Y. 2019, \mnras, 485, 648,
  \dodoi{10.1093/mnras/stz261}

\bibitem[{{Qiu} {et~al.}(2020){Qiu}, {Shannon}, {Farah}, {Macquart}, {Deller},
  {Bannister}, {James}, {Flynn}, {Day}, {Bhandari}, \& {Murphy}}]{qiu20}
{Qiu}, H., {Shannon}, R.~M., {Farah}, W., {et~al.} 2020, \mnras, 497, 1382,
  \dodoi{10.1093/mnras/staa1916}

\bibitem[{{Rafiei-Ravandi} {et~al.}(2021){Rafiei-Ravandi}, {Smith}, {Li},
  {Masui}, {Josephy}, {Dobbs}, {Lang}, {Bhardwaj}, {Patel}, {Bandura},
  {Berger}, {Boyle}, {Brar}, {Cassanelli}, {Chawla}, {Dong}, {Fonseca},
  {Gaensler}, {Giri}, {Good}, {Halpern}, {Kaczmarek}, {Kaspi}, {Leung}, {Lin},
  {Mena-Parra}, {Meyers}, {Michilli}, {M{\"u}nchmeyer}, {Ng}, {Petroff},
  {Pleunis}, {Rahman}, {Sanghavi}, {Scholz}, {Shin}, {Stairs}, {Tendulkar},
  {Vanderlinde}, \& {Zwaniga}}]{rafiei21}
{Rafiei-Ravandi}, M., {Smith}, K.~M., {Li}, D., {et~al.} 2021, arXiv e-prints,
  arXiv:2106.04354.
\newblock \doarXiv{2106.04354}

\bibitem[{{Rickett} {et~al.}(2009){Rickett}, {Johnston}, {Tomlinson}, \&
  {Reynolds}}]{rickett09}
{Rickett}, B., {Johnston}, S., {Tomlinson}, T., \& {Reynolds}, J. 2009, \mnras,
  395, 1391, \dodoi{10.1111/j.1365-2966.2009.14471.x}

\bibitem[{{Schechter}(1976)}]{schechter76}
{Schechter}, P. 1976, \apj, 203, 297, \dodoi{10.1086/154079}

\bibitem[{{Schoen} {et~al.}(2021){Schoen}, {Leung}, {Masui}, {Michilli},
  {Chawla}, {Pearlman}, {Shin}, {Stock}, \& {CHIME/FRB
  Collaboration}}]{schoen21}
{Schoen}, E., {Leung}, C., {Masui}, K., {et~al.} 2021, Research Notes of the
  American Astronomical Society, 5, 271, \dodoi{10.3847/2515-5172/ac3af9}

\bibitem[{{Scholz} \& {Stephens}(1987)}]{scholz87}
{Scholz}, F.~W., \& {Stephens}, M.~A. 1987, Journal of the American Statistical
  Association, 82, 918, \dodoi{10.2307/2288805}

\bibitem[{{Seta} {et~al.}(2021){Seta}, {Rodrigues}, {Federrath}, \&
  {Hales}}]{seta21}
{Seta}, A., {Rodrigues}, L. F.~S., {Federrath}, C., \& {Hales}, C.~A. 2021,
  \apj, 907, 2, \dodoi{10.3847/1538-4357/abd2bb}

\bibitem[{Shaffer(1995)}]{shaffer95}
Shaffer, J.~P. 1995, Annual Review of Psychology, 46, 561,
  \dodoi{10.1146/annurev.ps.46.020195.003021}

\bibitem[{{Shannon} {et~al.}(2018){Shannon}, {Macquart}, {Bannister}, {Ekers},
  {James}, {Os{\l}owski}, {Qiu}, {Sammons}, {Hotan}, {Voronkov}, {Beresford},
  {Brothers}, {Brown}, {Bunton}, {Chippendale}, {Haskins}, {Leach},
  {Marquarding}, {McConnell}, {Pilawa}, {Sadler}, {Troup}, {Tuthill},
  {Whiting}, {Allison}, {Anderson}, {Bell}, {Collier}, {G{\"u}rkan}, {Heald},
  \& {Riseley}}]{shannon18}
{Shannon}, R.~M., {Macquart}, J.~P., {Bannister}, K.~W., {et~al.} 2018, \nat,
  562, 386, \dodoi{10.1038/s41586-018-0588-y}

\bibitem[{{Shen} {et~al.}(2003){Shen}, {Mo}, {White}, {Blanton}, {Kauffmann},
  {Voges}, {Brinkmann}, \& {Csabai}}]{shen03}
{Shen}, S., {Mo}, H.~J., {White}, S. D.~M., {et~al.} 2003, \mnras, 343, 978,
  \dodoi{10.1046/j.1365-8711.2003.06740.x}

\bibitem[{{Simha} {et~al.}(2020){Simha}, {Burchett}, {Prochaska}, {Chittidi},
  {Elek}, {Tejos}, {Jorgenson}, {Bannister}, {Bhandari}, {Day}, {Deller},
  {Forbes}, {Macquart}, {Ryder}, \& {Shannon}}]{simha20}
{Simha}, S., {Burchett}, J.~N., {Prochaska}, J.~X., {et~al.} 2020, \apj, 901,
  134, \dodoi{10.3847/1538-4357/abafc3}

\bibitem[{{Spitler} {et~al.}(2016){Spitler}, {Scholz}, {Hessels}, {Bogdanov},
  {Brazier}, {Camilo}, {Chatterjee}, {Cordes}, {Crawford}, {Deneva}, {Ferdman},
  {Freire}, {Kaspi}, {Lazarus}, {Lynch}, {Madsen}, {McLaughlin}, {Patel},
  {Ransom}, {Seymour}, {Stairs}, {Stappers}, {van Leeuwen}, \&
  {Zhu}}]{spitler16}
{Spitler}, L.~G., {Scholz}, P., {Hessels}, J.~W.~T., {et~al.} 2016, \nat, 531,
  202, \dodoi{10.1038/nature17168}

\bibitem[{{Sun} {et~al.}(2015){Sun}, {Zhang}, \& {Li}}]{sun15}
{Sun}, H., {Zhang}, B., \& {Li}, Z. 2015, \apj, 812, 33,
  \dodoi{10.1088/0004-637X/812/1/33}

\bibitem[{{Tendulkar} {et~al.}(2017){Tendulkar}, {Bassa}, {Cordes}, {Bower},
  {Law}, {Chatterjee}, {Adams}, {Bogdanov}, {Burke-Spolaor}, {Butler},
  {Demorest}, {Hessels}, {Kaspi}, {Lazio}, {Maddox}, {Marcote}, {McLaughlin},
  {Paragi}, {Ransom}, {Scholz}, {Seymour}, {Spitler}, {van Langevelde}, \&
  {Wharton}}]{tendulkar17}
{Tendulkar}, S.~P., {Bassa}, C.~G., {Cordes}, J.~M., {et~al.} 2017, \apjl, 834,
  L7, \dodoi{10.3847/2041-8213/834/2/L7}

\bibitem[{{Totani}(2013)}]{totani13}
{Totani}, T. 2013, \pasj, 65, L12, \dodoi{10.1093/pasj/65.5.L12}

\bibitem[{{van den Bergh}(1999)}]{vandenbergh99}
{van den Bergh}, S. 1999, \aapr, 9, 273, \dodoi{10.1007/s001590050019}

\bibitem[{{Vedantham} \& {Phinney}(2019)}]{vedantham19}
{Vedantham}, H.~K., \& {Phinney}, E.~S. 2019, \mnras, 483, 971,
  \dodoi{10.1093/mnras/sty2948}

\bibitem[{{Wang} {et~al.}(2016){Wang}, {Yang}, {Wu}, {Dai}, \& {Wang}}]{wang16}
{Wang}, J.-S., {Yang}, Y.-P., {Wu}, X.-F., {Dai}, Z.-G., \& {Wang}, F.-Y. 2016,
  \apjl, 822, L7, \dodoi{10.3847/2041-8205/822/1/L7}

\bibitem[{{Xu} \& {Han}(2015)}]{xu15}
{Xu}, J., \& {Han}, J.~L. 2015, Research in Astronomy and Astrophysics, 15,
  1629, \dodoi{10.1088/1674-4527/15/10/002}

\bibitem[{{Xu} \& {Zhang}(2020)}]{xu20}
{Xu}, S., \& {Zhang}, B. 2020, \apjl, 898, L48,
  \dodoi{10.3847/2041-8213/aba760}

\bibitem[{{Yang} {et~al.}(2017){Yang}, {Luo}, {Li}, \& {Zhang}}]{yang17}
{Yang}, Y.-P., {Luo}, R., {Li}, Z., \& {Zhang}, B. 2017, \apjl, 839, L25,
  \dodoi{10.3847/2041-8213/aa6c2e}

\bibitem[{{Yao} {et~al.}(2017){Yao}, {Manchester}, \& {Wang}}]{yao17}
{Yao}, J.~M., {Manchester}, R.~N., \& {Wang}, N. 2017, \apj, 835, 29,
  \dodoi{10.3847/1538-4357/835/1/29}

\bibitem[{{Zheng} {et~al.}(2014){Zheng}, {Ofek}, {Kulkarni}, {Neill}, \&
  {Juric}}]{zheng14}
{Zheng}, Z., {Ofek}, E.~O., {Kulkarni}, S.~R., {Neill}, J.~D., \& {Juric}, M.
  2014, \apj, 797, 71, \dodoi{10.1088/0004-637X/797/1/71}

\bibitem[{{Zhu} {et~al.}(2018){Zhu}, {Feng}, \& {Zhang}}]{zhu18}
{Zhu}, W., {Feng}, L.-L., \& {Zhang}, F. 2018, \apj, 865, 147,
  \dodoi{10.3847/1538-4357/aadbb0}

\end{thebibliography}
\bibliographystyle{aasjournal}

%\listofchanges

\end{document}